\documentclass[preprint]{aastex}
\usepackage{pdfpages} 

\let\captionbox\relax

\usepackage{hyperref}

\def\deg{\ifmmode {^\circ}\else {$^\circ$}\fi}
\def\degree{\ifmmode {^\circ}\else {$^\circ$}\fi}
\def\mum{\ifmmode {\rm \,\mu {\rm m}}\else $\rm \,\mu {\rm m}$\fi}
\def\micron{\ifmmode {\rm \,\mu {\rm m}}\else $\rm \,\mu {\rm m}$\fi}
\def\arcsec{\ifmmode ^{\prime \prime}\else $^{\prime \prime}$\fi}

\def\inch{\ifmmode ^{\prime \prime}\else $^{\prime \prime}$\fi}
\def\msun{\ifmmode {M_{\odot}}\else $M_{\odot}$\fi}
\def\msunpery{\ifmmode {M_{\odot}\,{\rm yr}^{-1}}\else $M_{\odot}\,{\rm yr}^{-1}$\fi}
\def\lsun{\ifmmode {\rm L_{\odot}}\else $\rm L_{\odot}$\fi}
\def\mstar{\ifmmode {\rm M_{\star}}\else $\rm M_{\star}$\fi}
\def\lstar{\ifmmode {\rm L_{\star}}\else $\rm L_{\star}$\fi}
\def\md{\ifmmode {\rm M_d}\else $\rm M_d$\fi}
\def\ld{\ifmmode {\rm L_d}\else $\rm L_d$\fi}
\def\mearth{\ifmmode {\rm M_{\oplus}}\else $\rm M_{\oplus}$\fi}
\def\qdstar{\ifmmode Q_D^\star\else $Q_D^\star$\fi}
\def\kms{\ifmmode {\rm \,km\,s^{-1}}\else $\rm \,km\,s^{-1}$\fi}
\def\ms{\ifmmode {\rm m~s^{-1}}\else $\rm m~s^{-1}$\fi}
\def\mesc{\ifmmode m_{esc}\else $m_{esc}$\fi}
\def\rmin{\ifmmode r_{min}\else $r_{min}$\fi}
\def\rmax{\ifmmode r_{max}\else $r_{max}$\fi}
\def\mmin{\ifmmode m_{min}\else $m_{min}$\fi}
\def\mmax{\ifmmode m_{max}\else $m_{max}$\fi}
\def\rmind{\ifmmode r_{min,d}\else $r_{min,d}$\fi}
\def\rmaxd{\ifmmode r_{max,d}\else $r_{max,d}$\fi}
\def\mmaxd{\ifmmode m_{max,d}\else $m_{max,d}$\fi}
\def\vrad{\ifmmode v_{rad}\else $v_{rad}$\fi}
\def\qz{\ifmmode q_{0}\else $q_{0}$\fi}
\def\qi{\ifmmode q_{i}\else $q_{i}$\fi}
\def\ql{\ifmmode q_{l}\else $q_{l}$\fi}
\def\qs{\ifmmode q_{s}\else $q_{s}$\fi}
\def\rbrk{\ifmmode r_{brk}\else $r_{brk}$\fi}
\def\rdamp{\ifmmode r_{damp}\else $r_{damp}$\fi}
\def\ain{\ifmmode a_{in}\else $a_{in}$\fi}
\def\aout{\ifmmode a_{out}\else $a_{out}$\fi}
\def\r0{\ifmmode r_{0}\else $r_{0}$\fi}
\def\m0{\ifmmode m_{0}\else $m_{0}$\fi}
\def\M0{\ifmmode M_{0}\else $M_{0}$\fi}
\def\xm{\ifmmode x_{m}\else $x_{m}$\fi}
\def\gyr{\ifmmode {\rm g~yr^{-1}}\else ${\rm g~yr^{-1}}$\fi}
\def\icm{\ifmmode {\rm cm^{-1}}\else ${\rm cm^{-1}}$\fi}
\def\cms{\ifmmode {\rm cm~s^{-1}}\else ${\rm cm~s^{-1}}$\fi}
\def\gcms{\ifmmode {\rm g~cm^{-2}}\else $\rm g~cm^{-2}$\fi}
\def\gcmc{\ifmmode {\rm g~cm^{-3}}\else $\rm g~cm^{-3}$\fi}
\def\pcm{\ifmmode {\rm \,cm^{-1}}\else $\rm \,cm^{-1}$\fi}
\def\psqcm{\ifmmode {\rm \,cm^{-2}}\else $\rm \,cm^{-2}$\fi}
\def\pccm{\ifmmode {\rm \,cm^{-3}}\else $\rm \,cm^{-3}$\fi}
\def\water{\ifmmode {\rm H_2O}\else $\rm H_2O$\fi}
\def\cotwo{\ifmmode {\rm CO_2}\else $\rm CO_2$\fi}
\def\hm{\ifmmode {\rm H_2}\else $\rm H_2$\fi}
\def\acetylene{\ifmmode {\rm C_2H_2}\else $\rm C_2H_2$\fi}
\def\water{\ifmmode {\rm H_2O}\else $\rm H_2O$\fi}
\def\ammonia{\ifmmode {\rm NH_3}\else $\rm NH_3$\fi}

\graphicspath{{./}{figures/}}

\shorttitle{MIR Spectroscopy of GV Tau N}
\shortauthors{Najita et al.}

\begin{document}

\title{\bf High-Resolution Mid-Infrared Spectroscopy of GV Tau N: 
Surface Accretion and Detection of NH$_3$  in a Young Protoplanetary Disk}

\author{Joan R. Najita}
\affil{NSF's NOIRLab, 
950 N.\ Cherry Avenue, Tucson, AZ 85719, USA} 

\author{John S. Carr}
\affil{Department of Astronomy,  
University of Maryland, 
College Park, MD 20742, USA}

\author{Sean D. Brittain}
\affil{Clemson University, 
118 Kinard Laboratory, 
Clemson, SC 29634, USA}

\author{John H. Lacy}
\affil{Department of Astronomy, 
University of Texas at Austin, 
Austin, TX 78712, USA} 

\author{Matthew J. Richter}
\affil{Physics Department, 
University of California at Davis, Davis, CA 95616, USA } 


\author{Greg W. Doppmann}
\affil{W.\ M.\ Keck Observatory, 
65-1120 Mamalahoa Hwy., Kamuela, HI 96743, USA}

\begin{abstract}
Physical processes that redistribute or remove 
angular momentum from protoplanetary disks can 
drive mass accretion onto the star and 
affect the outcome of planet formation. 
Despite ubiquitous evidence 
that protoplanetary disks are engaged in accretion,
the process(es) responsible remain unclear. 
Here we present evidence for redshifted molecular 
absorption in the spectrum of a Class I source   
that indicates rapid inflow at the disk surface. 
High resolution mid-infrared spectroscopy of GV Tau N 
reveals a rich absorption 
spectrum of individual lines of 
\acetylene, HCN, \ammonia, and \water. 
From the properties of the molecular absorption, 
we can infer that it carries a significant accretion rate 
$\dot M_{\rm abs} \sim 10^{-8} -10^{-7}\msunpery,$
comparable to the stellar accretion rates of active T Tauri stars. 
Thus we may be observing disk accretion in action. 
The results may provide observational evidence 
for supersonic ``surface accretion flows,'' which have been found 
in MHD simulations of magnetized disks.  
The observed spectra also represent the first detection of \ammonia\ in 
the planet formation region of a protoplanetary disk. 
With \ammonia\ only comparable in abundance to HCN, 
it cannot be a major missing reservoir of nitrogen. 
If, as expected, the dominant nitrogen reservoir in 
inner disks is instead N$_2$, its high volatility 
would make it difficult to incorporate into forming 
planets, which may help to explain the low nitrogen 
content of the bulk Earth.

\end{abstract}


\section{Introduction} \label{sec:intro}

Stars form surrounded by disks, the repositories of 
excess angular momentum inherited from the molecular 
cloud that cannot be contained within the star alone. 
Physical processes that redistribute or remove 
disk angular momentum can drive mass accretion 
onto the star and 
affect the outcome of planet formation; 
e.g.,  giant planets that form in the disk may be swept inward 
along with the accretion flow.
Although it is as yet unclear what physical mechanism(s) 
generate accretion flows, the process(es) must be 
robust: T Tauri stars (i.e., young stars surrounded by disks) 
commonly show excess UV emission produced by mass 
flows that reach the stellar surface 
(Hartmann et al.\ 2016). 

One clue to the nature of disk accretion 
comes from the sizes of gaseous disks. 
Whereas disks in the embedded ``Class I'' phase of pre-main-sequence
evolution are modest in size (typically $< 100$\,au in radius), gas
disks in the more evolved `‘Class II'’ phase, when infall from the
molecular cloud has ceased, span a range of sizes including many
much larger than 300\,au (Najita \& Bergin 2018; Ansdell et al.\
2018). The existence of such large disks is remarkable given the
multiple processes that can act to reduce disk sizes (e.g., FUV
photoevaporation, companion formation, disk winds). Their existence
suggests that internal angular momentum redistribution plays a large
enough role in driving accretion that a significant fraction of
disks grow substantially in size from the Class I to Class II phases
(Najita \& Bergin 2018). 
However there have been few observational clues to date 
as to the nature of the process that redistributes the 
angular momentum. 
Disk spectral line diagnostics offer the opportunity to 
probe the dynamics of disks for potential new insights. 

Spectral features in the mid-infrared (MIR) 
offer the opportunity to study the 
dynamics and chemistry of planet formation 
environments. 
Class II (T Tauri) disks commonly reveal a 
MIR spectrum that is rich in emission 
from water and organic molecules 
(\cotwo, HCN, \acetylene; Carr \& Najita 2008, 2011; Salyk et al.\ 2011). 
The emitting areas, line profiles, and 
thermal-chemical properties of the emitting gas 
are consistent with emission from an irradiated disk 
atmosphere within a few au of the star, i.e., the 
terrestrial planet region of the disk 
(e.g., Najita \& \'Ad\'amkovics 2017; Najita et al.\ 2018).
In contrast to the frequent detection of MIR molecular 
emission, MIR molecular {\it absorption} 
is detected only rarely from young stars at low 
spectral resolution. Observations with the {\it Spitzer 
Space Telescope} detected molecular absorption 
features of HCN, \acetylene, and \cotwo\ from 
IRS 46 (Lahuis et al.\ 2006) and   
GV Tau (e.g., Bast et al.\ 2013), 
and \cotwo\ from DG Tau B 
(Kruger et al.\ 2011; Pontoppidan et al.\ 2008). 
The similar molecular species and gas temperatures detected in 
emission and absorption suggested that the absorption features 
also arise in a disk atmosphere but viewed close to 
edge on (Bast et al.\ 2013; Doppmann et al.\ 2008). 

Here we report high resolution MIR spectroscopy of the 
Class I source GV Tau N, which reveals a rich molecular absorption 
spectrum of individual lines of 
\acetylene\ and HCN, as well as \water\ and 
the first detection of \ammonia\ in 
the planet formation region of a protoplanetary disk. 
The resolved line profiles, which have a significant 
redshifted component, may provide observational evidence 
for a new accretion mechanism: 
supersonic ``surface accretion flows,'' which have been found 
in MHD simulations of magnetized disks (Zhu \& Stone 2018; 
see also Stone \& Norman 1994; Beckwith et al.\ 2009; 
Guilet \& Ogilvie 2012, 2013).

A young stellar binary in the Taurus molecular cloud (d=140 pc), 
GV Tau (a.k.a. Haro 6--10, IRAS 04263+2426)
comprises a bright optical source, GV Tau S, and an infrared companion
located 1.2\arcsec\ away, GV Tau N, which dominates the MIR flux of  
the system (e.g., Leinert \& Haas 1989; Sheehan \& Eisner 2014).
While GV Tau has the strongly rising 2--25\,\micron\ continuum 
characteristic
of Class I sources (e.g., Furlan et al.\ 2008), its relatively weak
gas and dust emission compared to other low-mass embedded YSOs
suggests that the system lacks a significant envelope 
(Hogerheijde et al.\ 1998). That property and the poorly 
defined molecular outflow structure of the system 
(Hogerheijde et al.\ 1998) suggest that it is a
more evolved Class I source. Spatially resolved millimeter imaging
further suggests modest solid disk masses for both GV Tau N and S
(Sheehan \& Eisner 2014) and small dust disk sizes 
($\sim 15$ au; Guilloteau et al.\ 2011).

In addition to the molecular absorption bands of HCN, \acetylene,
and \cotwo\ detected with {\it Spitzer} (Bast et al.\ 2013),  
the spectrum of GV Tau N 
also shows individual near-infrared (NIR) absorption lines of 
HCN, \acetylene, CH$_4$, and CO 
(Gibb et al.\ 2007; Doppmann et al.\ 2008; Gibb \& Horne 2013; 
Davis et al.\ 2015). 
While CO absorption has been detected toward both
Class I and Class II sources at high spectral resolution 
(Brittain et al.\ 2005; Rettig et al.\ 2006; Horne et al.\ 2012; 
Kruger et al.\ 2011; Smith et al.\ 2015, Lee et al.\ 2016), 
organic absorption features are rarely seen. 
Here we report a 
high spectral resolution observation of 
the MIR spectrum of GV Tau N. 
The observations are described in Section 2, 
and the detected features are analyzed in Section 3. 
In Section 4, we discuss the origin of the absorption features 
and the possible support they provide for the existence of 
surface accretion flows. We also comment on the detection of 
\ammonia\ and its implications for our understanding of the 
nitrogen reservoir in disks.

\section{TEXES Observations} \label{sec:obs}

We observed GV Tau N with TEXES, the Texas Echelon-cross-Echelle
spectrograph (Lacy et al.\ 2002), on the Gemini-North 8m telescope
during observing campaigns in November 2006 and October 2007.  We
used the high resolution, cross-dispersed mode for all observations
with a slit width of 0.54$^{\prime\prime}$.  The spectral resolving
power in this instrumental configuration is R$\approx$100,000.  We
nodded the object on the slit every $\sim$10 seconds to facilitate
background subtraction and removal of night-sky emission.  TEXES
does not use the chopping secondary mirror.
During the first visit to GV Tau on (UT) 19 November 2006, we observed both GV Tau S and GV Tau N to confirm that the mid-IR flux originated almost entirely with GV Tau N.  Subsequent to that observation, we concentrated solely on GV Tau N.  

The spectral coverage in a single setting with TEXES is roughly
0.5\%\ of the central wavenumber. Therefore, we targeted specific
molecular transitions in each setting based on the telluric atmosphere,
the source Doppler shift, and a desire to cover a range of lower
state levels and molecules. We used the night-sky emission features
to evaluate and adjust the spectral setting to avoid desired features
falling into gaps in the coverage; for $\lambda \geq 11\ \mu m$,
the TEXES spectral orders are larger than the detector.

The standard observation sequence with TEXES includes observing an
ambient temperature blackbody roughly every 7 minutes as a flatfield.
We also observed a telluric calibrator, either a hot star or a
featureless continuum source, immediately before or after GV Tau N.
The telluric calibrator helps to reduce the impact of atmospheric
absorption and to flatten the continuum beyond what division by 
the blackbody alone provides.  A log of the observations, 
including the telluric calibrator used, is given in 
Table 1.

The data were reduced using the standard TEXES pipeline.  The
pipeline removes spikes, rectifies the cross-dispersed echellograms,
flatfields the data, aligns and combines nod pairs, optimally
extracts the spectrum of the continuum object, and does a wavelength
calibration using a night-sky emission line.  We subsequently divided
the GV Tau N spectra by a scaled version of the telluric calibrator
spectrum to remove residual atmospheric lines.

The wavelength scale was refined using telluric absorption lines
in the spectrum of GV Tau N. For each setting, a dispersion function
was fit to all measureable telluric lines in each order, using the
IRAF task ecidentify.\footnote{IRAF was distributed by the National 
Optical Astronomy Observatory, which was managed by the Association 
of Universities for Research in Astronomy (AURA) under a cooperative 
agreement with the National Science Foundation.} 
The fits had an average rms residual of
$0.15\,\kms$ and an average maximum residual of $0.3\,\kms$.  Two
observations have larger uncertainties. The signal-to-noise for 
the $814\,\pcm$
setting from 2007 was too low to use telluric lines in the GV Tau N 
spectrum.  Instead, the dispersion function was fit to lines in the
telluric standard (rms $0.06\,\kms$), and this solution was applied
to GV Tau N.  Three measurable telluric lines in the GV Tau N spectrum
were used to assign a zero-point (wavelength shift) uncertainty of
$0.5\,\kms$.  The $932\,\pcm$ setting from 2007 only contains three
telluric lines; hence an independent dispersion solution was not
possible. In this case, the pipeline wavelength scale was retained,
with an assumed uncertainty of $1\,\kms$.

Model telluric transmission spectra were utilized for a few of the
settings.  Because the S/N for the $780\,\pcm$ setting was degraded
due to insufficient S/N in the telluric calibrator, corrections for
atmospheric lines were made using the software package TERRASPEC
(Bender 2010). TERRASPEC was also used to improve the correction
for telluric water lines in the 767 and $807\,\pcm$ settings.  For
all of the observations, additional continuum normalization was
carried out for each order of interest by fitting a low-order
polynomial to the continuum outside the intervals of absorption
from known molecular transitions in GV Tau N.

\section{Analysis}

In the TEXES spectra we detect numerous absorption lines of \ammonia,
HCN, and \acetylene, a few lines of \water\ in absorption, and \hm\
in emission (Table 2).  
The \hm\ emission from GV Tau was previously reported
in Bitner et al.\ (2008) and is not discussed further here. 
Most of the HCN and \acetylene\ lines
were measured in 2006, with additional lines measured in 2007. The
unexpected detection in 2006 of numerous \ammonia\ absorption lines
was followed up in 2007 with grating settings that probed lower
energy \ammonia\ transitions. Weak absorption lines of water were 
also detected in the 2007 spectra. 
The entire set of pipeline-reduced spectra for GV Tau N are shown in 
Figure A1.  
The positions and identifications of the detected lines are indicated.

\begin{figure}[htb!]
\includegraphics[width=1.\linewidth,trim={0.0in 0.5in 0.0in 0.5in},clip]{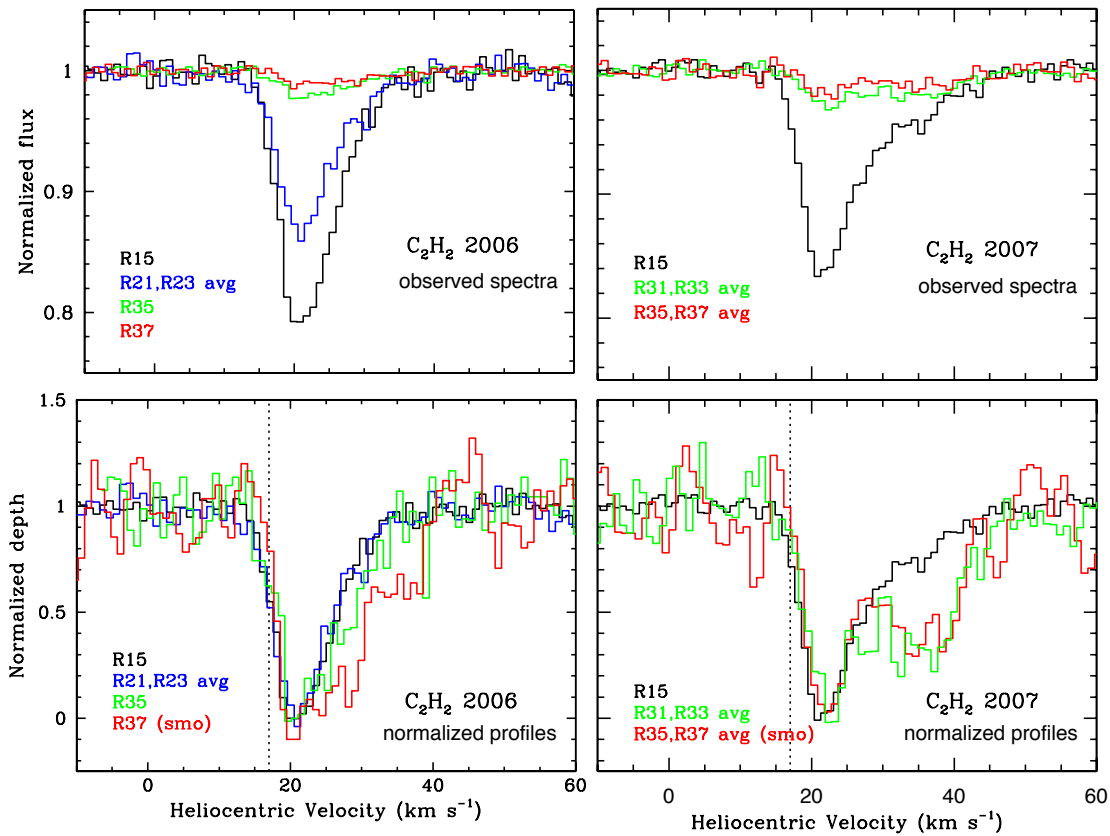}
\caption{\scriptsize \acetylene\ spectra (top) and 
normalized line profiles (bottom)
in 2006 (left) and 2007 (right).
For reference, in the bottom panels, the vertical lines mark
the velocity of the molecular cloud core from Hogerheijde et al.\ (1998).
In the bottom panels, the profiles are scaled to extend from 
0 at the bottom of the absorption to 1 in the continuum. 
The lines have an absorption core at $v_{\rm helio} \simeq 20\kms$
and a red wing extending to $\sim 40 \kms.$
In both epochs,
the strength of the red wing relative to the core is greater in
the higher energy lines (above R23) than the lower energy lines.
}
\label{fig:fig1}
\end{figure}

While the \water\ lines are pure rotational transitions within the
ground vibrational state, the \ammonia, HCN, and \acetylene\
absorption lines are predominantly 
ro-vibrational transitions out of the ground
vibrational state to a higher vibrational state; the \acetylene\
$\nu_4$+$\nu_5$ -- $\nu_4$ line is an exception (see section 3.1).

The HCN $\nu_2$ vibrational mode is the bending mode.  
The $\nu_5$
mode of \acetylene\ (HCCH) is the symmetric bending mode, in which the two
H atoms vibrate together, and the $\nu_4$ mode is the antisymmetric
bending mode.  In the R5 transition, for example, the rotational transition
is J=5 to J=6, Q5 is J=5 to J=5, and P5 is J=5 to J=4.  For \acetylene, 
the H atom spin statistics cause the odd-J (ortho) lines to be three times
as strong as the even-J lines, and Q-branch lines are approximately
twice as strong as P- and R-branch lines.  Since HCN and \acetylene\ are
linear molecules in their ground vibrational states, only the quantum
number J is required to label their rotational states, although the
excited bending modes are split into e and f states depending on
the relative orientation of the rotational axis and the bending
motion.  

For \ammonia, the $\nu_2$ vibrational mode is the `umbrella' mode in which
the three H atoms vibrate together.  The ground and $\nu_2$ states are
split by inversion splitting, with lines from the symmetric and
anti-symmetric states labelled s and a.  For this symmetric top
molecule the rotational state is described by two quantum numbers,
J (the total angular momentum) and K (the angular momentum about
the symmetry axis).  Line strengths depend on both J and K, but
typically states with K=3n are twice as strong as those with K=3n$\pm$1.
J can change in a radiative transition, but K does not.  Transitions
are labelled by the lower state J and K.

\subsection{HCN and \acetylene}

We detect a total of 8 lines of the HCN $\nu_2$ band, 
14 lines of the \acetylene\ $\nu_5$ band, and 
one line from the \acetylene\ $\nu_4$+$\nu_5$ -- $\nu_4$ band
(i.e., absorption out of the $\nu_4$ vibrational level;
Figure A1).
In 2006, we detected 12 \acetylene\ lines between R5 and R37 
and 6 HCN lines between R10 and R34. 
In 2007, we detected 6 \acetylene\ lines between R15 and R37, as well as  
three HCN transitions (R17, R18, R32) with limits on two additional 
HCN transitions (R33 and R34). 
The lower energy levels of the observed lines range from 
$E_\ell < 100$\,K to $\sim 1600$\,K. 

The absorption features range in strength from $\sim 1$\% deep at
high excitation down to a maximum of $\sim 20$\% deep at 
intermediate-$J$ values (Fig.~1 and Fig.~2).
Figure 1 shows representative line profiles of \acetylene\ 
over a range of excitation for 2006 (left) and 2007 (right), both 
as observed (top) and with the absorption features scaled to the same
depth (bottom) in order to compare relative velocity profiles.
In the bottom panels, vertical lines indicate 
the velocity of the gaseous envelope surrounding GV Tau
(dashed line; $v_{\rm helio} = 17.3\pm 0.5 \kms$; 
or $v_{\rm LSR} = 7.0\,\kms$; Hogerheijde et al.\ 1998).
The MIR molecular absorption is clearly redshifted with 
respect to the molecular cloud velocity.

\begin{figure}[htb!]
  \centering
\includegraphics[width=0.5\linewidth,trim={0.0in 0.5in 0.0in 0.5in},clip]{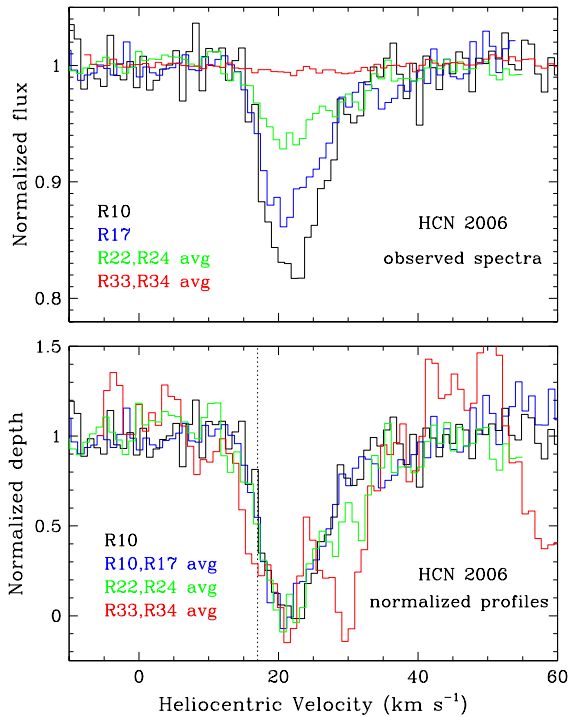}
  \caption{\scriptsize HCN spectra (top) and 
normalized line profiles (bottom) in 2006.
Like the \acetylene\ lines, higher energy HCN lines show
a red wing that is stronger relative to the core
than lower energy HCN lines.}
  \label{fig:fig2}
\end{figure}

\begin{figure}[htb!]
  \centering
\includegraphics[width=0.5\linewidth,trim={0.0in 0.5in 0.0in 0.5in},clip]{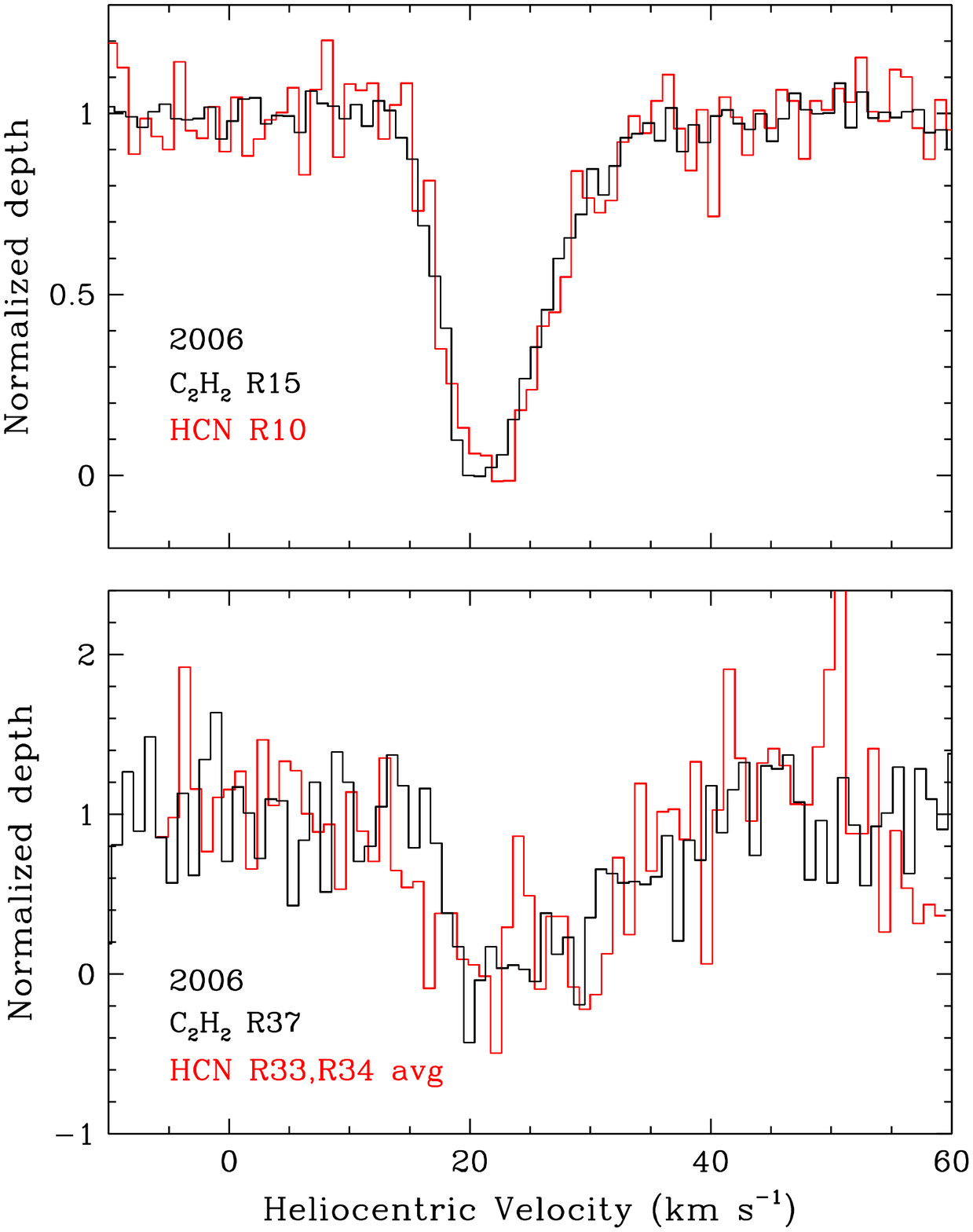}
  \caption{\scriptsize The HCN and \acetylene\ lines showed similar profiles at
low-$J$ (top) and high-$J$ (bottom) in 2006 and 2007 (not shown),
suggesting that the HCN and \acetylene\ absorption arise in the same gas along
the line of sight.}
  \label{fig:fig3}
\end{figure}

Figure 2 shows the HCN spectra and normalized profiles 
from the 2006 data; the HCN profiles from 2007 are similar. 
Figure 3 compares the \acetylene\ and HCN profiles 
for low (top) and high (bottom) excitation lines.
The \acetylene\ and HCN lines have nearly identical 
velocity profiles.

At low-$J$,
the HCN and \acetylene\ lines profiles both have 
a ``single dip'' core absorption component 
centered at 21--22 \kms\,
with a FWHM of $\sim 8-10 \kms$,
and a red wing extending
to $\sim 40 \kms$ (Figs.~1, 2, 3). 
When scaled to the same depth, the \acetylene\ and 
HCN profiles show a trend in which the red wing becomes stronger
(relative to the core)
at high $J$ (Fig.~1 and Fig.~2), 
i.e., the high velocity gas is more highly excited or is
more optically thick. 
While low and intermediate rotational levels (e.g., R15 of \acetylene)
tend to show a wing that smoothly decreases in strength to the red,
higher excitation lines show more velocity structure. For example, the
\acetylene\ R37 line could be fit with velocity components at 21, 28,
and 36~\kms. 
The velocity structure of the \ammonia\ (\S 3.2) and 
\water\ (\S S3.3) lines is similar to that seen in the 
HCN and \acetylene\ lines, suggesting that the
molecular absorption is composed of multiple velocity components.

\begin{figure}[htb!]
        \centering
\includegraphics[width=0.5\linewidth,trim={0.0in 0.5in 0.0in 0.5in},clip]{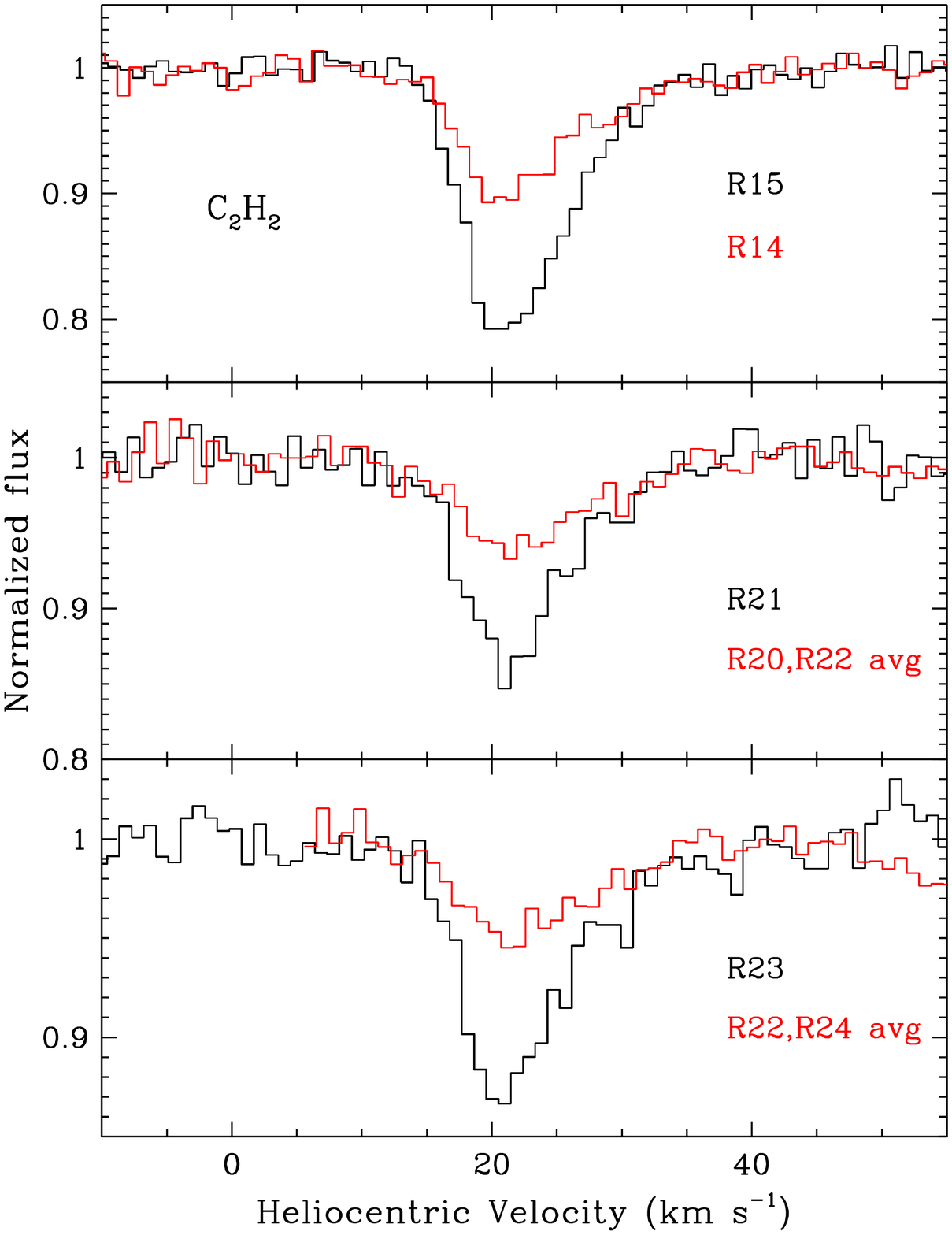}
\caption{\scriptsize Comparison of \acetylene\ ortho and para lines in 2006.
If the absorbing gas is optically thin, the odd-J lines should be
approximately 3 times stronger than the adjacent even-J lines.
The more modest ratio of the absorption depths implies that the
absorbing gas is marginally optically thick.
The relative absorption depths approach unity in the red wing,
indicating larger optical depths than in the line core. }
\label{fig:fig4}
\end{figure}

If the absorbing gas is optically thin, the odd-$J$ \acetylene\ 
lines should be approximately 3 times
stronger than the adjacent even-$J$ lines.
For several of the stronger \acetylene\ lines observed 
(R14 vs.\ R15; R21 vs.\ the average of R20 and R22; 
R23 vs.\ the average of R22 and R24), the ratio of the absorption 
depths of the ortho and para lines is lower, $\sim 2$ in the 
absorption core (Fig.~4), indicating that 
the odd-$J$ lines have optical depths of $\sim 1$ or greater.
The relative depths of the neighboring ortho-para lines 
also differ between the core and the wing.
The ratio of the absorption depths declines from $\sim 2$ within the
absorption core to $\sim 1$ in the red wing beyond 30\,\kms\ 
(Fig.~4),
indicating that the even-$J$ \acetylene\ lines are also optically thick
in the high velocity gas.
Interestingly, the observed depths of the \acetylene\ lines ($< 20$\%)
are far less than expected for optically thick lines. 
This suggests that the line strengths are significantly diluted,
or an intrinsically narrow line is highly broadened by macroscopic
motions, or some combination of these two effects.

\begin{figure}[htb!]
  \centering
\includegraphics[width=0.5\linewidth,trim={0.0in 0.5in 0.0in 0.5in},clip]{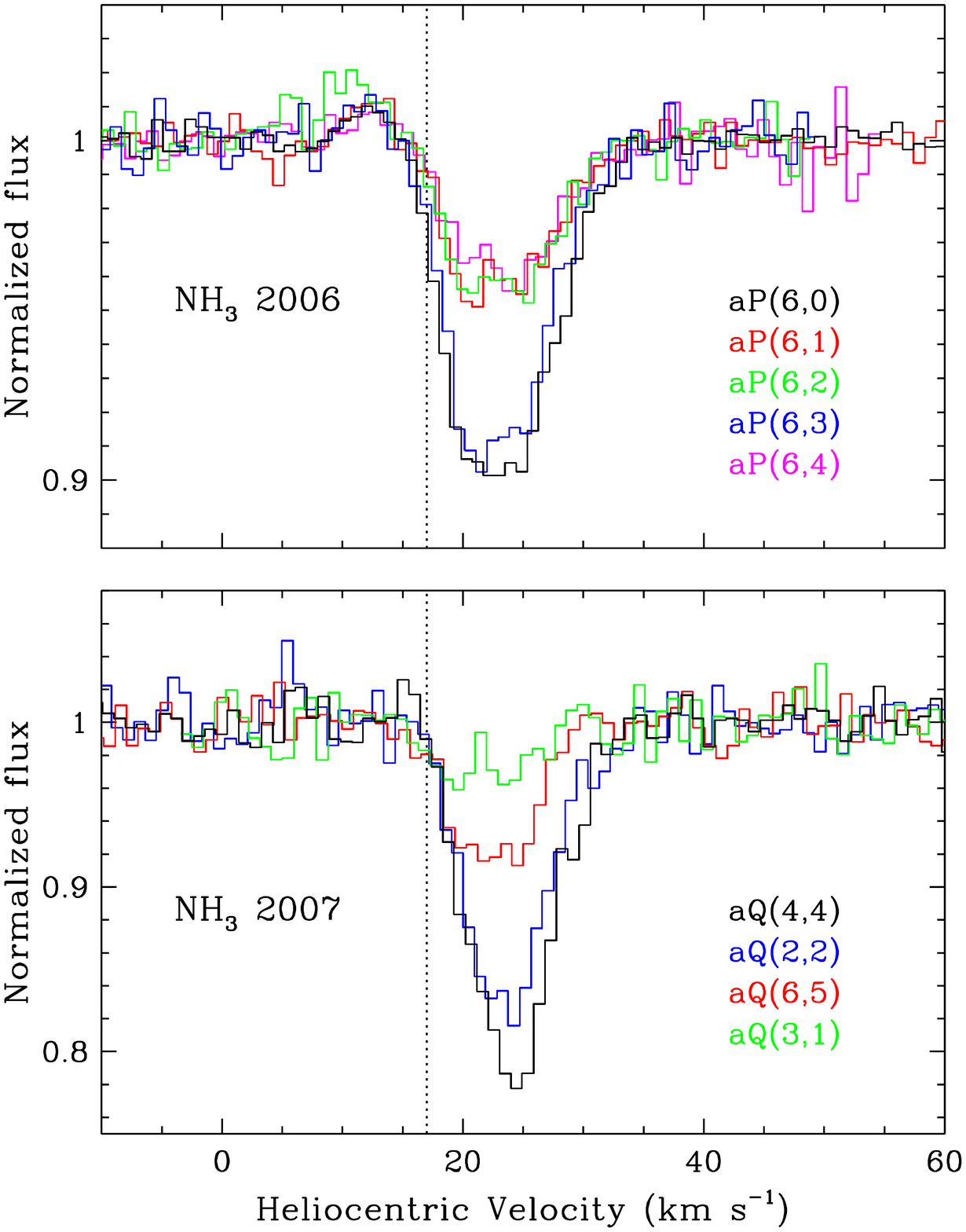}
  \caption{\scriptsize Spectra of \ammonia\ absorption lines.
The top panel shows the aP(6,K) transitions observed in 2006.
The two strongest lines have twice the statistical weight
of the others but similar profiles.
The bottom panel shows a selection of Q-branch lines of various strengths observed
in 2007. The weaker lines have a more flat-bottomed profile and a bluer
velocity centroid.
}
  \label{fig:fig5}
\end{figure}

\subsection{\ammonia}

We detect a total of 34 lines of the \ammonia\ $\nu_2$ band (Figure A1): 
12 transitions in 2006 with lower energy levels from 
$358 \pcm$ to $888 \pcm$, and  
27 transitions in 2007 with lower energy levels from 
$17 \pcm$ to $1581 \pcm$. 
The \ammonia\ absorption strengths are similar to those of the \acetylene\ and
HCN lines and range in strength from $\sim 1$\% 
deep for the weakest measured lines
to a maximum of $\sim 30$\% deep 
for the aQ(3,3) transition measured in 2007.
Example spectra for individual \ammonia\ lines are shown in 
Figure 5 
for several P-branch lines from 2006 and Q-branch lines from 2007.

\begin{figure}[htb!]
  \centering
\includegraphics[width=0.5\linewidth,trim={0.0in 0.3in 0.0in 0.5in},clip]{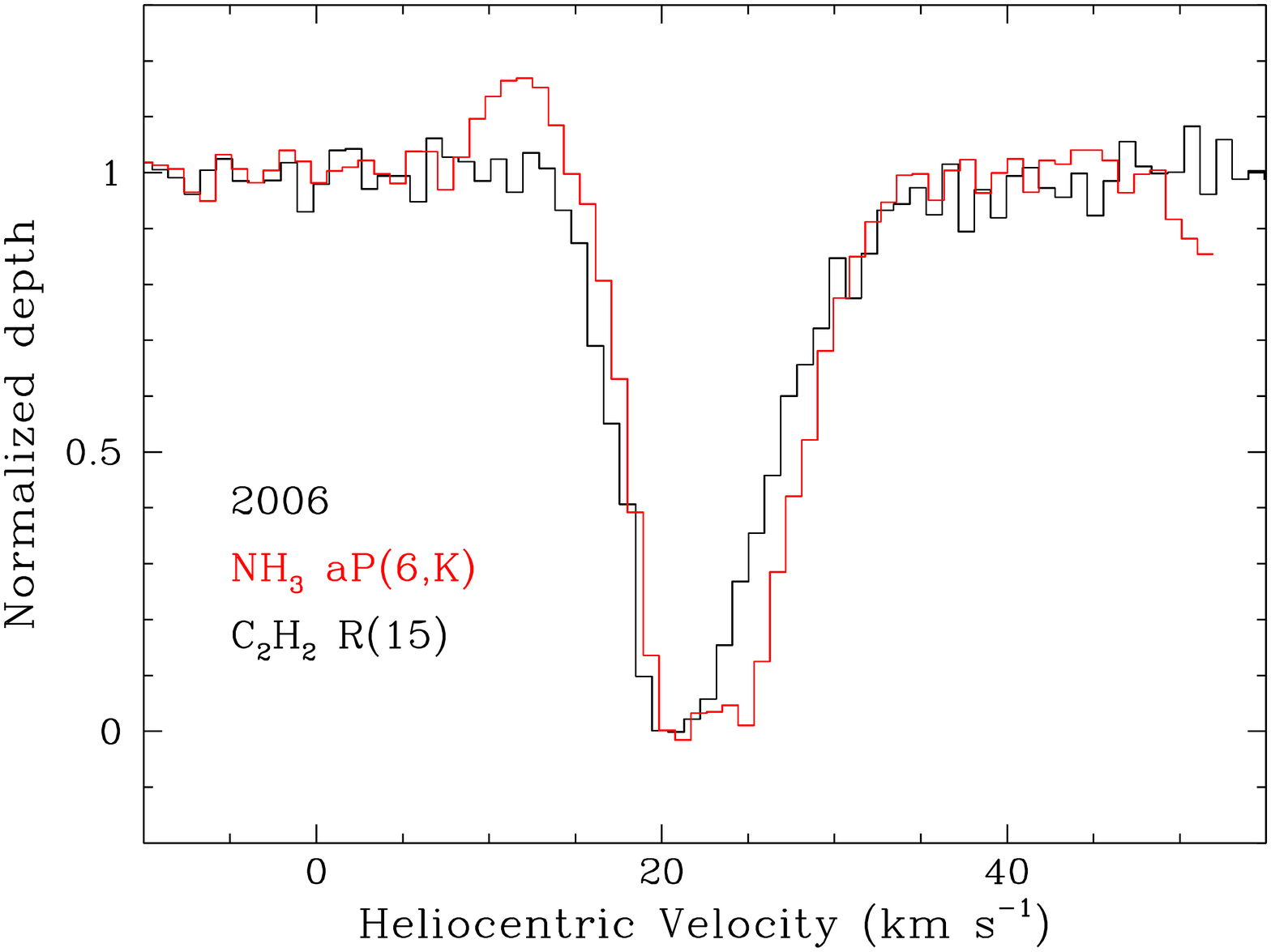}
  \caption{\scriptsize Comparison of the 2006 \ammonia\ aP(6,K) 
average profile with that of the \acetylene\ R15 
reveals similar profiles, suggesting that the \ammonia\ and \acetylene\
arise in the gas along similar lines of sight.
}
  \label{fig:fig6}
\end{figure}

The profiles of the \ammonia\ P-branch lines are roughly 
similar to the \acetylene\ and HCN profiles, although there 
are differences in detail.
Figure 6 compares 
the \acetylene\ R15 line ($E_\ell = 282\,\icm$) with   
the average of four aP(6,K) \ammonia\ lines from 2006
($E_\ell =384-417\,\icm$). 
Like the HCN and \acetylene\ line profiles discussed in the previous
section, the P-branch \ammonia\ line profiles also have
an absorption core at $\sim 22\kms$ and a red wing extending to 
$\sim 40 \kms$.
However, the shape of the \ammonia\ core is more square bottomed,
with a small shift in the velocity of the blue edge.
In addition, the aP(6,K) \ammonia\ lines from 2006 show 
weak emission blueward of the absorption. 
A similar emission component is not seen in the line profiles of 
the other molecules, nor in the profiles of other \ammonia\ lines 
(including the aP(6,K) lines from 2007); however, the latter 
could be due to lower signal-to-noise of the other spectra.
Similar line profiles are predicted in radiative transfer models of 
disk atmospheres with modest radial flows 
(see Section 4.2).

\begin{figure}[htb!]
  \centering
\includegraphics[width=0.5\linewidth,trim={0.0in 0.3in 0.0in 0.5in},clip]{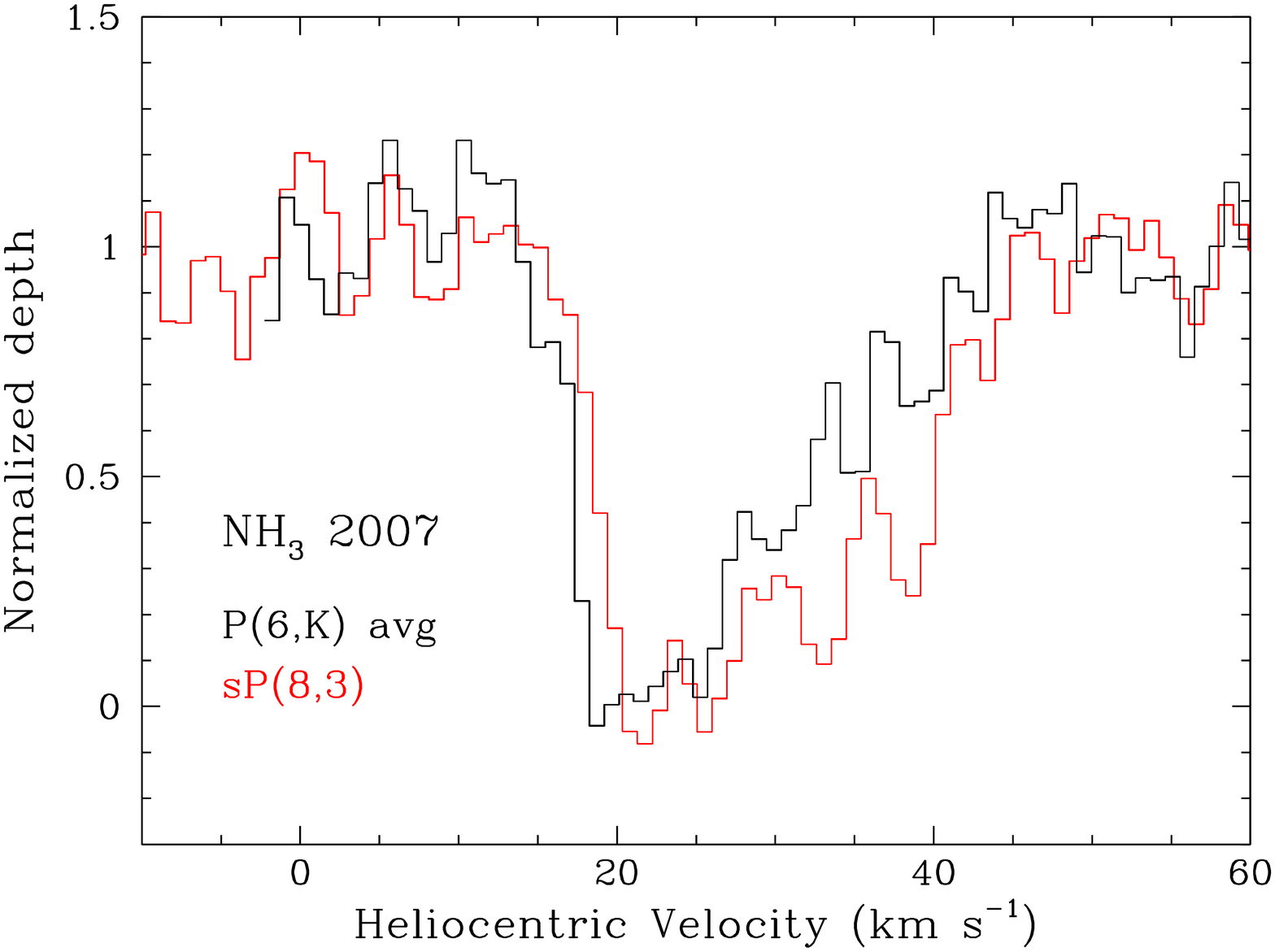}
  \caption{\scriptsize Comparison of the \ammonia\
aP(6,K) average profile ($E_\ell \sim 400\,\pcm$)
and the sP(8,3) profile ($E_\ell$ = 679\,\pcm) observed in 2007
shows that, like the \acetylene\ and HCN lines,
the high velocity wing is stronger in the higher energy transitions.}
  \label{fig:fig7}
\end{figure}

\begin{figure}[htb!]
  \centering
\includegraphics[width=0.5\linewidth,trim={0.0in 0.3in 0.0in 0.5in},clip]{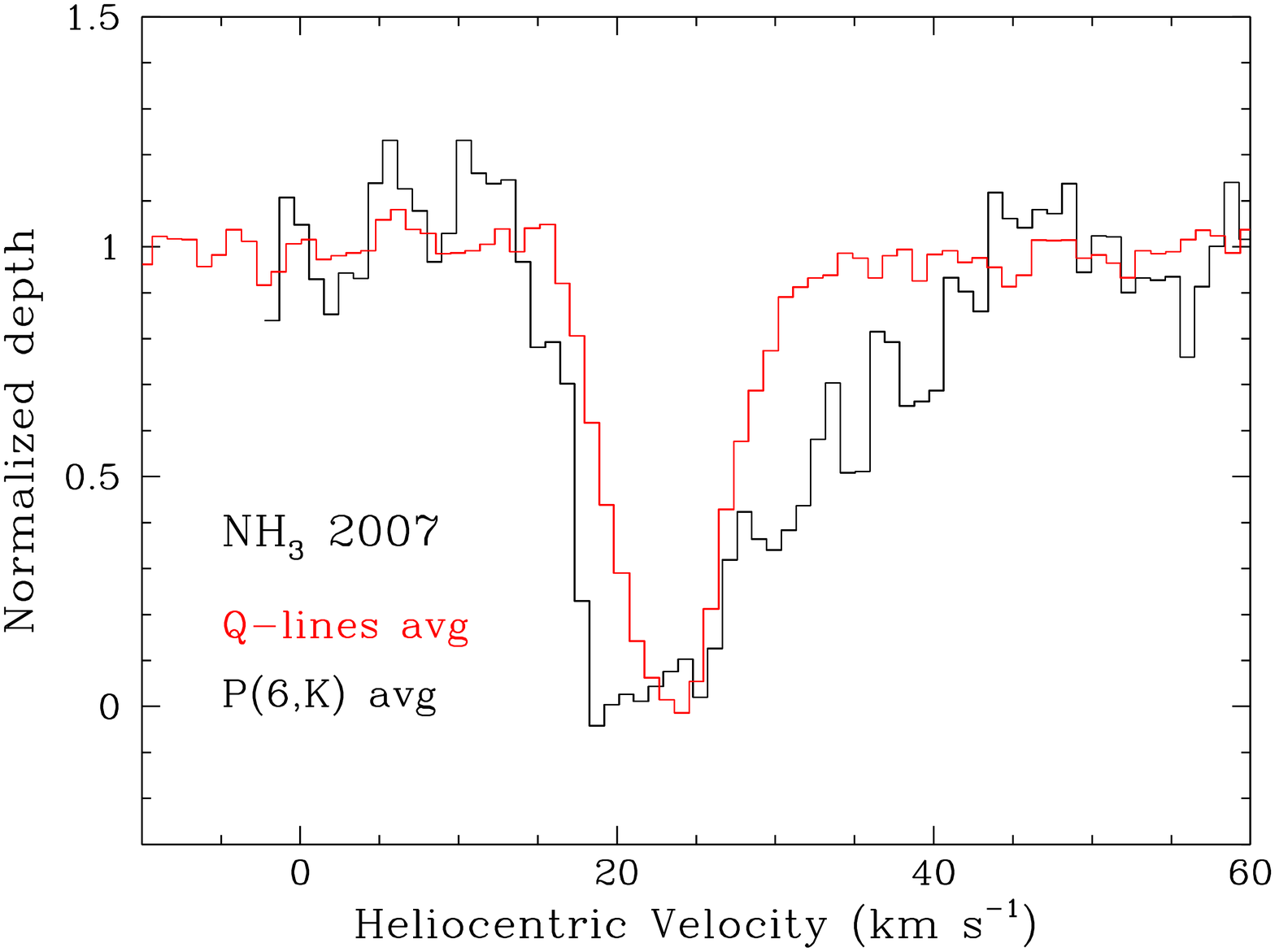}
  \caption{\scriptsize Comparison of \ammonia\ profiles for an average of
seven Q-branch lines and four aP(6,K) lines
of similar energy in 2007.
The line shapes differ markedly, with the Q-branch \ammonia\ lines
lacking much of a red wing.
}
  \label{fig:fig8}
\end{figure}

Similar to \acetylene\ and HCN, the higher velocity red wing of the
\ammonia\ profiles becomes stronger, relative to the core, for
higher excitation lines.
Figure 7 compares the average 2007 profile of
the aP(6,K) lines ($E_\ell = 384-417\,\icm$) 
with that of the sP(8,3) line ($E_\ell = 679\,\icm$).
The higher energy line clearly shows 
enhanced absorption at higher velocities in the red wing. 

Perhaps surprisingly,  
the Q-branch lines show the $\sim 22\,\kms$
core absorption component (FWHM $\sim 8\,\kms$) 
but {\it lack} the red wing.
Figure 8 compares the average 2007 line profiles of 
the Q-branch ($E_\ell = 17-423\,\icm$)  
and aP(6,K) lines.  
The profile shapes 
differ markedly, with the higher velocity absorption in the red wing 
absent in the Q-branch lines. 
The average Q-branch profile is sharper and centered more
to the red than the flat-bottomed P-branch lines.  However, the
Q-branch profiles also vary with line strength (see Fig. 5, bottom),
with the weaker Q-branch lines closer in shape and velocity 
to those of the aP(6,K) lines.
The absence of the red wing in the Q-branch lines 
may have something to do with the fact that 
the Q-branch transitions overlap 
the 10\,\micron\ silicate absorption feature, 
while the measured P-branch lines fall between 12 and 13\,\micron.
Detailed radiative transfer models that include these opacity 
differences may be able to help us understand the origin of 
the differences in the profiles.

\subsection{\water}

We detected 4 high excitation lines of \water\ in the 2007 spectra 
(Figure A1), 
with lower energy levels ranging from 2631\,\pcm\ to 3211\,\pcm. 
All of the lines are very weak, $\sim$ 2--3\% deep.
In contrast to the \acetylene, HCN, and \ammonia\ lines, which 
have a prominent absorption core at $\sim 22\,\kms$,
the \water\ absorption is deepest at $\sim 37 \kms$.

\begin{figure}[htb!]
  \centering
\includegraphics[width=0.5\linewidth,trim={0.0in 0.5in 0.0in 0.5in},clip]{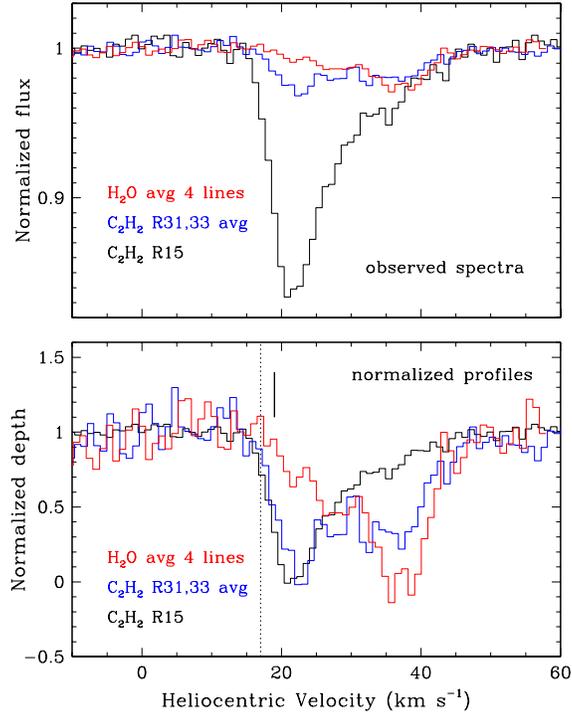}
  \caption{\scriptsize Comparison of spectra (top) and 
normalized line profiles (bottom) of
water lines in 2007 to \acetylene\ lines.
The short solid vertical line marks the position of the telluric water absorption.
The vertical dotted line marks the velocity of the molecular cloud core.
}
  \label{fig:fig9}
\end{figure}

\begin{figure}[htb!]
  \centering
\includegraphics[width=0.5\linewidth,trim={0.0in 0.3in 0.0in 0.5in},clip]{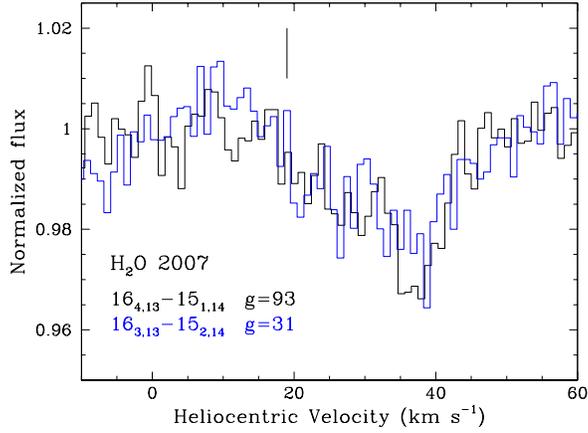}
  \caption{\scriptsize The similar strengths of ortho and para water lines in
2007 reveals that the absorption is optically thick.
The short solid vertical line marks the position of the telluric water absorption.}
  \label{fig:fig10}
\end{figure}

Figure 9 compares the average spectrum and profile of the four
detected \water\ lines 
with spectra and profiles of high-excitation (average of R31 and R33) 
and low-excitation (R15) lines of \acetylene\ in the 2007 data. 
The comparison shows that the \water\ absorption profile is strongest
in the far red wing of the low-$J$ \acetylene\ absorption profile.
The \water\ absorption is very weak in the region of the  
the 22~\kms\ component
and intermediate in strength at $\sim 28 \kms.$ 
The profile of the higher energy \acetylene\ line, 
whose overall strength is weak, like the \water\ absorption, 
is intermediate between the two profiles:
it has the same velocity components as \water\ (at roughly
22\,\kms, 28\,\kms, and 36\,\kms) but with more equal depths.

Figure 10 compares the profiles of the 
16$_{3,13}$--15$_{2,14}$ and 
16$_{4,13}$--15$_{1,14}$ water lines, which are 
an ortho-para pair;
they have nearly identical Einstein A-values and lower level energies
(Table 2)  
but differ by a factor of 3 in their statistical weights. 
The absorption strengths
of the two transitions are essentially the same, which shows that
the absorption is optically thick. At the same time, the absorption
depths are very small, which implies that the absorption is highly
diluted, either because the absorber covers a small fraction (filling
factor) of the background continuum and/or an intrinsically narrow,
deeper absorption feature is broadened by macroscopic motions. 
A similar, 
but less extreme, result was found for the \acetylene\ absorption.

Table 2 lists 4 the detected water lines as well as an 
additional 4 undetected water lines that were used to constrain 
the properties of the water absorption.

\subsection{Variability}

There are some differences in the absorption strength and 
velocity profiles of the lines observed in 2006 and 2007.
Figure 11 compares the 2006 and 2007 spectra of \acetylene\ for
the R15 line and the average of two high excitation lines. 
In 2007 the core of the R15 line decreased in strength, while the
red wing has increased. In the high excitation lines, the wing also
increased in strength in 2007 while the 22~\kms\ core remained 
about the same.
The same behavior is seen for HCN. The increase in the depth of the
red wing relative to the core is common to all transitions and to
all three molecules, including the P-branch \ammonia\ lines
observed in both epochs. 

A more subtle change is a small redward shift in the blue edge of the
lower excitation \acetylene\ and HCN lines, resulting in a change of the
line centroid from 22 to 23~\kms. These velocities differ from the
average velocity of 19~\kms\ that Doppmann et al.\ (2008) found for
the 3\,\micron\ HCN lines, in spectra with 13~\kms\ resolution that
did not resolve the line profiles. Hence, velocity variations may be
common, but the TEXES data show that this can be produced by changes
in the velocity structure of the absorption, rather than just a simple
velocity shift of the entire profile.

\begin{figure}[htb!]
  \centering
\includegraphics[width=0.5\linewidth,trim={0.0in 0.5in 0.0in 0.5in},clip]{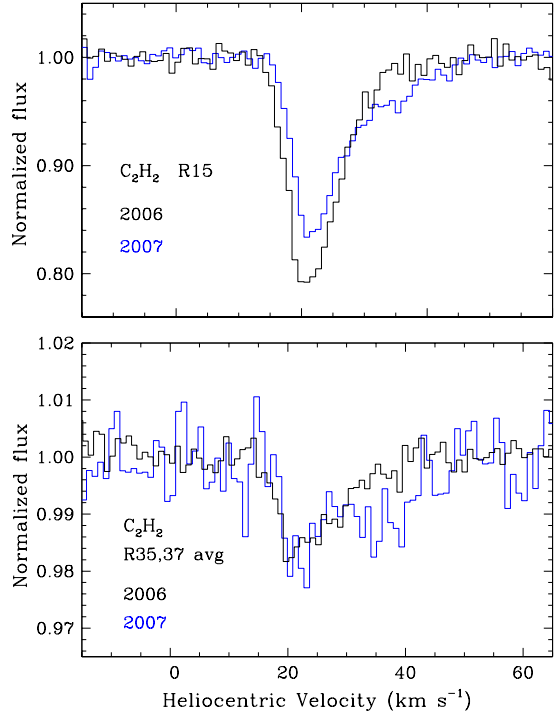}
  \caption{\scriptsize Comparison of profiles of high-$J$ (top) and
low-$J$ (bottom) profiles of
\acetylene\ lines in 2006 and 2007.
}
  \label{fig:fig11}
\end{figure}

\subsection{Equivalent Widths and Absorption Properties}

The equivalent widths (EWs) of the absorption features were used to
infer the temperatures and column densities of the absorbing molecules.
Because there are differences in absorption strengths and profiles
between 2006 and 2007, the EWs measured in the two epochs were analyzed
separately.

As noted in previous sections, the ``core'' and ``wing'' components
of the line profiles show different behaviors as a function of 
line excitation energy, both in the observed 
ortho-para ratio of \acetylene\ absorption depths 
and in the interesting absence of a red wing in the Q-branch 
\ammonia\ lines. 
We therefore measured and analyzed the absorption as that of 
two velocity components, simply defined: 
one redward of $\sim 26 \kms$ and the other blueward of the same velocity.
A possible alternative would have been to decompose the profile 
into two Gaussians.
However, because the profiles are not strictly Gaussian, 
and they in fact appear to be
composed of multiple components with possibly complex behavior, 
we adopted the simpler approach, with the goal of capturing the essence of
the difference between the lower and higher velocity absorption.

The measured EWs for the two velocity components
and their combined EW are given in Tables 3 and 4 
for all lines measured 
in 2006 and 2007, respectively. The uncertainties include both the 
spectral pixel-to-pixel noise, which comes from the continuum
signal-to-noise, 
and an
estimate of the uncertainty in the continuum placement. 
The average line depth over the velocity interval is also listed 
for both components.
In analyzing the detected \water\ absorption, we found it useful 
to also include upper limits on a few additional \water\ lines 
that could become detectable at very high column density and 
lower temperatures. These lines are listed in Table 4 along 
with their 2-$\sigma$ upper limits on EW for the high-velocity component.

In the case of a uniform absorbing medium that lies in front of an 
opaque continuum source,  
the absorption equivalent width, which is a simple function of the line
optical depth $\tau_\nu,$ 
can be diluted by 
unobscured continuum emission (e.g., if the absorber does not 
completely cover the background MIR continuum source) 
and/or by emission from the absorber.
That is, if the emission from a background source $B_{\nu,b}(T_b)$ with
temperature $T_b$ is absorbed by a medium with 
opacity $\tau_\nu$ and source function $S_{\nu,a}$, 
the emergent intensity is
\begin{equation}
I_\nu = B_{\nu,b}\,e^{-\tau_\nu} + 
\left(1-e^{-\tau_\nu}\right) S_{\nu,a}
\end{equation}
at a frequency $\nu$ in the line and $I_\nu = B_{\nu,b}$
in the neighboring continuum.
The fractional intensity decrement in the line relative to the
continuum is then
\begin{equation}
\left(B_{\nu,b}-I_\nu\right)/B_{\nu,b} =
\left(1-e^{-\tau_\nu}\right)\left(1-S_{\nu,a}/B_{\nu,b}\right),
\end{equation}
and the equivalent width 
\begin{equation}
EW = f_c \int\left(1-I_\nu/B_{\nu,b}\right) d\nu\,
= f_d \int\left(1-e^{-\tau_\nu}\right)d\nu, 
\end{equation}
where the quantity 
\begin{equation}
f_d = f_c \left(1-S_{\nu,a}/B_{\nu,b}\right)
\end{equation}
combines the effects of the covering fraction of
the absorber $f_c$
and the average dilution by emission from the absorber across the line
$f_e=\left(1-S_{\nu,a}/B_{\nu,b}\right).$
In LTE, $S_\nu=B_\nu(T)$ where $T$ is the temperature of 
the absorbing medium. However, as a result of the high 
critical density for vibrational LTE, the vibrational temperature 
$T_{\rm vib}$ may be less than $T,$ and both $S_\nu$ and $T_{\rm vib}$ 
can be affected by the radiation field.

Here we model the absorbing gas as a uniform slab. 
The line optical depths are determined by the gas temperature ($T$),
and the ratio of the column density of the molecule ($N$) 
to the intrinsic line width, 
which includes thermal and microturbulent components. 
If the absorption is approximated as 
a Gaussian
line profile 
with FWHM $\Delta v$, the line center optical depth is
\begin{equation}
\tau_0 = {g_u A \bar\nu^{3} \over 8\pi} {e^{-E_\ell/kT}\over Q(T)}
{N\over 1.0645\,\Delta v}\left(1 - e^{-hc\bar\nu/kT}\right)
\end{equation}
where $\bar\nu$ is the wavenumber of the transition,
$g_u$ is the multiplicity of the upper state, 
$E_\ell$ is the energy of the transition, 
$Q$ is the partition function, 
and the factor 1.0645 relates the product of 
the FWHM and peak height of a Gaussian to its area.
All of the molecular parameters and data for the partition 
functions are from HITRAN (Rothman et al\ 2013; Fischer et al.\ 2003). 
We assume that the lower energy levels are in LTE.

For each molecule and velocity component, we plot a curve-of-growth (COG) as
ln(EW) 
versus the logarithm of a quantity proportional to $\tau_0$ and then 
determine the model parameters that minimize $\chi^2$. 
The first two parameters, $T$ and $N/\Delta v$,  
determine the line optical depths and the shape of the COG; 
equivalently, these two parameters set the relative EWs of the lines.
The third parameter in this approach, $f_d\Delta v$, is a scaling factor that
best matches the absolute EWs for all lines. From these three parameters,
the value of $f_dN$ automatically follows.

Note that $f_d$ and $\Delta v$ are not independently determined, although it
is possible to place physical limits on $\Delta v$ (and hence $f_d$).
If all lines are optically thin (all points fall on the 
linear part of the COG), $N/\Delta v$ is not constrained and 
the only determined parameters are $T$ and $f_dN$. 
The results from the COG analysis are given in Tables 5 and 6.
The parameter uncertainties represent 1-$\sigma$, i.e., 
a 68\% confidence interval, where the joint confidence region for
the parameters is the space enclosed by the surface of constant
$\chi^2$ that corresponds to the confidence level and the number
of free parameters. Note that there is correlation between parameters,
such that the upper bound on temperature generally corresponds to
lower column density and larger intrinsic line width, and visa
versa.  The uncertainties depend on the number of lines used, the
signal-to-noise of the EWs, and the range in optical depth and lower
energy level covered by the measured transitions. 

In Figure 12, example COGs are plotted for both the 
low- and high-velocity components of \acetylene\ from 2006. 
The COG analysis shows quantitatively the conclusion reached 
from visual examination of the \acetylene\ spectra;
the high-velocity (HV) component (the red wing) 
has a higher optical depth than
the low-velocity (LV) component (the 22\,\kms\ core), 
as indicated by the greater curvature in its COG.
The line with the highest opacity is the R15, which has an optical depth
of 2 for the LV and 6 for the HV component.
Remarkably, the temperatures of the HV and LV components
are about the same, $\sim 450$\,K.
Thus, our finding that higher rotational transitions are more
prominent in the HV component is not due to a higher
temperature for the high-velocity gas but to greater optical depth
(larger $N/\Delta v$).

\begin{figure}[htb!]
  \centering
    \includegraphics[width=0.45\linewidth,trim={0.0in 0.5in 0.0in 0.5in},clip]{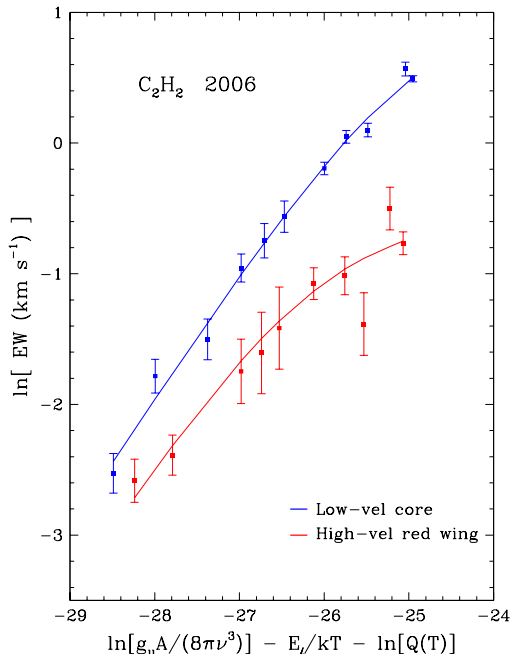}
  \caption{\scriptsize Curves of growth for the low-velocity (blue points)
and high-velocity (red points) components of \acetylene\ in 2006.
The solid lines are the best-fit model curves of growth.
}
  \label{fig:fig12}
\end{figure}

Despite its higher optical depth, the EWs and absorption depths of 
the HV  component are smaller than those 
of the LV component, indicating a larger dilution factor.
Because the LV and HV components share 
the same temperature (Tables 5, 6), the contribution of line emission 
by the warm absorber to the value of $f_d$ 
(through the quantity $f_e$ in eq.~1) 
is the same for both components.
Thus, the greater dilution of the HV component 
(i.e., smaller $f_d$) compared to the LV 
component indicates that 
the covering fraction of the background continuum 
is smaller for the high-velocity gas 
than the low-velocity gas.\footnote{Small values of 
$f_d$ reflect small covering fractions 
rather than significant emission from the absorbing gas. 
The vibrational temperature of the absorber $T_{\rm vib}$ 
and 
the temperature of 
the background continuum $T_b$ would have to be very similar 
for emission to have a significant impact.
In LTE, $T_{\rm vib}$ is the same as the rotational 
temperature of the absorber, which is here $\sim 450$\,K.
For $T_{\rm vib}$=450\,K, 
reducing $f_e$ to 0.16--0.21, the range of dilution factors inferred 
for the LV component (Tables 5 and 6), requires a background continuum 
temperature of $T_b$=480--490\,K. 
Such similar foreground and background temperatures seem implausible; 
with temperatures that similar, the foreground gas is as likely 
to produce emission as absorption. 
}

For \water, 
the absorption is dominated by the HV component 
(Section 3.3); the observed redward velocity of the 
absorption shifts it out of 
the telluric water absorption, enabling its detection. 
The $20\,\kms$ component is clearly detected in the average spectrum 
(Fig.~9), but it is too weak in individual lines to be useful in 
determining the absorption parameters. 
Therefore, the analysis was restricted to the HV component.
Furthermore, because all four of the detected lines are optically thick, 
we are unable to constrain the absorption properties using these
lines alone. We therefore also included in the analysis the upper limits on several 
additional water lines covered by our spectra. 
Because these additional lines, shown in Table 4, would
become strong enough to detect at very high column density and/or
low temperature, including their upper limits 
places an upper bound on the water column density. 
The 2-$\sigma$ upper limits were included in the $\chi^2$ fit using the approach
discussed in Sawicki (2012). The best-fit COG for the \water\ absorption is shown
in Figure 13.
The resulting analysis yields the best-fit properties shown in Table 6.

\begin{figure}[htb!]
  \centering
    \includegraphics[width=0.45\linewidth,trim={0.0in 0.5in 0.0in 0.5in},clip]{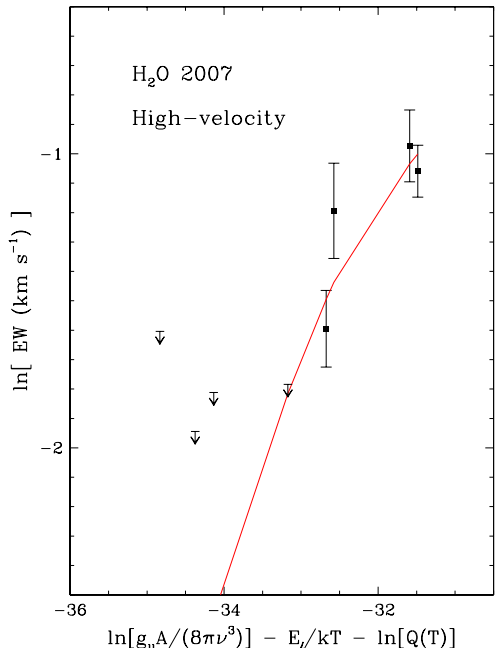}
  \caption{\scriptsize Curve of growth for \water. Detected water lines
(solid points) and upper limits (downward arrows) are
indicated along with the best fit (red line).
}
  \label{fig:fig13}
\end{figure}

Using the above procedure, 
we derived gas temperatures of $\sim 450$\,K for both velocity components
of all molecules, with one clear exception discussed below.  
For molecules other than water, 
the derived values for $N/\Delta v$ are in the range  
$10^{16} - 10^{17} \psqcm\,{\rm km^{-1}\,s}$,
with the value for the HV component consistently 3--7 times greater
than that for the LV component.
For the HV component of water, $N/\Delta v$ is much larger, 
$2.9\times 10^{19} \psqcm\,{\rm km^{-1}\,s}$.

Because radiative transitions between rotational levels of 
\acetylene\ are forbidden, collisions will control the 
rotational populations in the ground vibrational state. 
Therefore, LTE should be a good
approximation for \acetylene\ and the derived temperature a measure
of the gas kinetic temperature. For HCN, the critical densities for
our measured transitions range from $10^8-10^9\,\pccm$. The derived
excitation temperature is the same as that of \acetylene\ (for the
2006 data), indicating that the gas densities are at least this
high. The measured rotational levels for \ammonia\ have comparable
critical densities.
However, the critical densities for the measured water lines are higher,
$\sim 4\times 10^{10}\,\pccm.$ 
Models of disk atmospheres predict that warm molecules are present 
at high densities ($\gtrsim 10^{11}\,\pccm$), making LTE a reasonable 
assumption (e.g., Bruderer 2014; Najita \& \'Ad\'amkovics 2017). 

To determine the line-of-sight column density $N$ from the 
constrained quantity $N/\Delta v$ requires an assumption 
about the local line broadening $\Delta v.$ 
A lower limit on $N$ comes from assuming that the local line 
broadening is thermal, which for the observed molecules is 
$\sim 1\,\kms$ at 450\,K.
For an upper limit to $N,$ we can 
adopt for $\Delta v$ the observed width of the absorption component 
($\sim 8\,\kms$), which would be appropriate if the absorbing medium 
is highly turbulent. 
However, this is unlikely, because
the profiles of the R15 and R14 lines of \acetylene\ are identical across
the LV component, i.e., the ortho-para ratio and hence the optical depth
is nearly constant across the LV core of the lines.
If the absorption was due to a single intrinsically broad ($\sim 8\,\kms$) profile,
then the optical depth would be highest at the absorption minimum and decrease
away from line center, in contrast to the observed profiles.

More likely, the observed width of the absorption feature
is produced by multiple lines of sight through the rotating disk atmosphere
that pass through gas at different velocities.
In this case, the local line broadening $\Delta v$ is substantially less than
the full observed width, implying a smaller line-of-sight column density.
In Tables 5 and 6 we list values for $N$ for
an assumed width of $2\,\kms,$ a value between 
the approximate thermal width of hydrogen ($4.5\,\kms$) and  
that of water ($\sim 1\,\kms$)
at the temperature of the absorption (450\,K), 
implying some amount of microturbulent broadening for the observed molecules.
We interpret these values of $N$ in the next section.

Although we find a $\sim 450$\,K temperature for the 
velocity components of almost all molecules, 
we derive a much lower temperature for \ammonia\ in 2007 (250\,K)
than in 2006 (455\,K) for the LV component. 
Because the addition of the Q branch in the 2007 data adds a 
large number of lines, which cover a greater range in optical depth, 
the temperature from the COG is well constrained.
Setting the temperature to 250\,K for the 2006 data gives a extremely bad fit. 
One could be concerned that the Q branch probes different gas, given its
different behavior, i.e., the absence of the red wing in the Q-branch lines.
In Figure 14, we plot the COG for the 2007 LV component,
where the model COG is the best fit to the combined Q- and P-branch lines.
The P-branch lines are plotted with a different color, and follow the 
same trend as the Q-branch lines. In addition, fitting the
Q- and P-branch lines separately yields consistent temperatures.

\begin{figure}[htb!]
  \centering
    \includegraphics[width=0.45\linewidth,trim={0.0in 0.5in 0.0in 0.5in},clip]{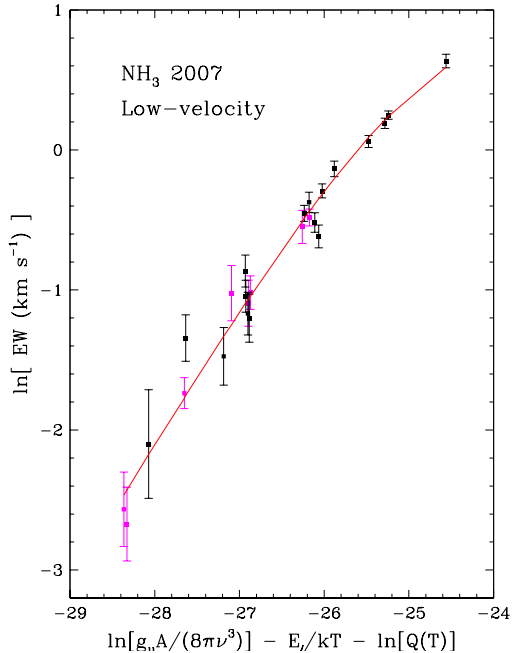}
  \caption{\scriptsize Curve of growth for the 2007 low-velocity
component of \ammonia. The Q-branch (black) and P-branch (pink)
lines follow the same trend and are fit with similar temperatures.
The solid line is the best-fit model curve of growth.
}
  \label{fig:fig14}
\end{figure}

{\it Comparison with Previous Results.} 
We can compare our results for HCN to the findings of Doppmann et al.\ (2008),
who measured the 3\,\micron\ absorption lines of HCN and \acetylene\ in GV Tau N
at a resolution of 13\,\kms\ using NIRSPEC on Keck.
They fit the $\nu_1$ HCN spectrum with
a temperature of 550\,K and an HCN column density of 
$1.5\times 10^{17}\psqcm$, using a microturbulent velocity
of 3\,\kms\ (FWHM = 5.0\,\kms).
This is equivalent to $N/\Delta v = 3\times 10^{16}\psqcm\,{\rm km^{-1}\,s}$.
Hence, their result is in reasonable agreement with the values we obtained
with the MIR lines for the LV component of HCN.

We can also compare the results from our spectrally resolved
molecular spectra with the analysis by Bast et al.\ (2013) of
the {\it Spitzer}/IRS spectrum of GV Tau. Our study differs
from that of Bast et al.\ in two important ways.
Firstly, the TEXES data spatially resolve the N and S components
of the binary, whereas Bast et al.\ did not account for
dilution by GV Tau S.
Our TEXES observations confirm that
the S component lacks MIR molecular absorption,
a result similar to that found
in $L$-band molecular spectroscopy of GV Tau N and S
(Doppmann et al.\ 2008; Gibb et al.\ 2007).
Secondly, the high spectral resolution of the TEXES data allows us
to resolve the profiles and to infer dilution of the line equivalent widths.

For comparison to Bast et al., we use the $f_d\,N$ values from our analysis
with $f_d$ = 1, i.e., the column densities we would derive without taking dilution
into account.
For \acetylene\ and HCN, our column densities are somewhat larger, by factors of 1--3,
than those reported by Bast et al. The temperature for HCN is in agreement,
but their temperature for \acetylene\ (720 K) is higher.
For \ammonia, our values of $f_d\,N \sim 2\times 10^{16}\psqcm$ are similar
to the upper limit reported by Bast et al.\ ($1.9\times 10^{16}\psqcm$ for 500\,K).
All of our derived column densities would be larger for $f_d < 1$.
Some of the observed differences may be due to variability, 
because the {\it Spitzer} and TEXES data were taken at different 
times. 

To summarize, 
the TEXES spectra reveal warm ($\sim 450\,K$), 
redshifted MIR molecular absorption in GV Tau N, 
with larger column densities than inferred in previous 
studies. 
The HV component of the absorption is more optically thick 
than the LV component, has a column density 3--6 times larger, 
and is more highly diluted (by a factor $\sim 5$) than 
the LV component 
due to a smaller covering fraction of the background 
MIR continuum.

\section{Discussion} \label{sec:dis}

\subsection{Origin of the Absorption} 

The working hypothesis in the literature
is that the molecular absorption in GV Tau N
occurs in gas in the disk atmosphere 
observed against the hotter dust continuum from smaller radii
in a disk viewed close to edge-on
(Gibb et al.\ 2007; Doppmann et al.\ 2008; Bast et al.\ 2013).
Here we discuss the origin of the absorbing gas
in light of our results on the temperature, column densities,
and line profiles of the molecular absorption. 

{\it Clues from Temperature and Column Density.}\quad 
We find that the absorbing gas is warm ($\sim 450$ K) 
and the detected molecular species show 
profiles that can be modeled as multiple velocity components 
that probe gas along the same  
lines of sight to the 12\,\micron\ continuum.
The rotational temperatures derived for the HCN, \acetylene, 
and \water\ absorption are similar to (but at the cool end of) the
excitation temperatures derived for the same molecules from the
MIR molecular emission observed from T Tauri disks 
(400--1000\,K; Carr \& Najita 2011; Salyk et al.\ 2011). 
The latter is thought to arise from the
heated disk atmosphere within an au of the star 
(e.g., Najita et al.\ 2011; Najita \& \'Ad\'amkovics 2017).

The relative column densities of the molecular species 
detected in absorption in GV Tau N
are also 
consistent with 
the ratios derived for the 
MIR molecular emission from T Tauri disks.  
For the high-velocity absorption component measured for 
GV Tau N in 2007, for which a
\water\ column density could be determined,
the column density ratios are
$N(\acetylene)/N(\water)$ = 0.002,
$N({\rm HCN})/N(\water)$ = 0.005,
and $N(\acetylene)/N({\rm HCN})$ = 0.4.
Figure 15 compares the GV Tau N column density 
ratios (blue star) with the emission column density 
ratios of T Tauri disks derived from 
high and low resolution spectra (squares). 
The GV Tau N values are similar to the ratios derived 
from T Tauri star emission spectra using slab models 
that make assumptions similar to those adopted here. 
The similar column density ratios are consistent with 
the interpretation that the GV Tau N absorption arises 
in a disk atmosphere. 

\begin{figure}[htb!]
  \centering
    \includegraphics[width=0.5\linewidth,trim={0.0in 0.0in 0.0in 0.0in},clip]{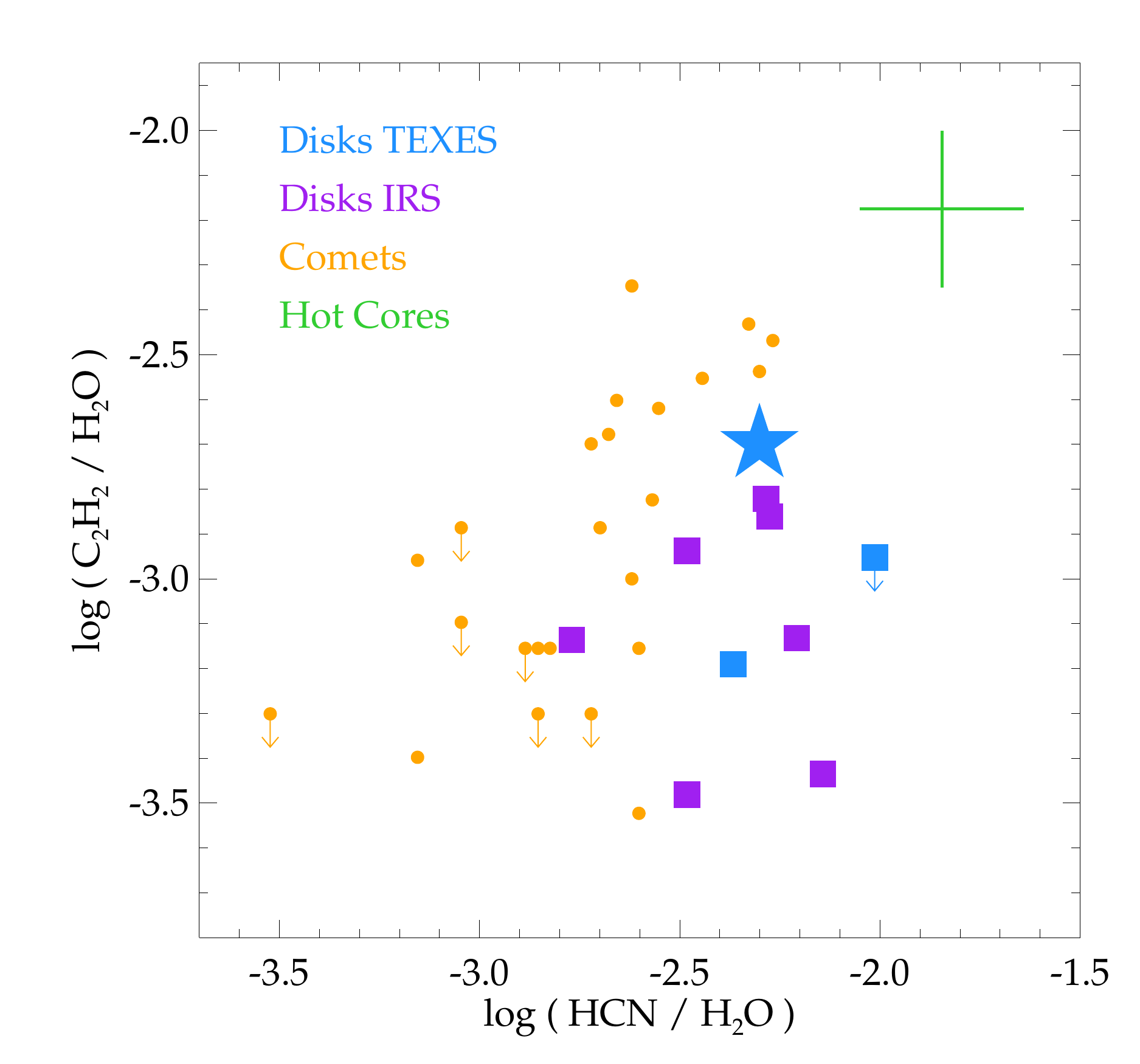}
  \caption{\scriptsize Molecular absorption column density ratios reported here
for GV Tau N (blue star) compared with
emission column density ratios from T Tauri disk spectra studied at
high spectral resolution with TEXES (blue squares; Najita et al.\ 2018)
and
low spectral resolution with {\it Spitzer}/IRS (purple squares;
Carr \& Najita 2011; Najita et al.\ 2018).
The estimates from {\it Spitzer} assume equal emitting areas for all molecules.
Also shown are the abundances of comets
(orange dots; Dello Russo et al.\ 2016),
and the average properties of hot cores (green cross;
see Carr \& Najita 2011 for details).
The GV Tau N absorption column density ratios are similar to
the emission column density ratios of T Tauri disks, consistent
with the interpretation that the absorption arises in a disk
atmosphere viewed at high inclination.
}
  \label{fig:fig15}
\end{figure}

At the same time, the absolute column densities of the MIR absorption
are much higher than the column densities seen in emission.
Here we compare specifically to results for AS205 N and DR Tau, two
high accretion rate disks viewed at low inclination. 
As the only sources whose HCN and \acetylene\ emission has been 
measured at high spectral resolution, their velocity-resolved 
spectra enable tighter constraints on 
the properties of the emitting gas (Najita et al.\ 2018). 
For a valid comparison (apples to apples), 
we assume thermal local linewidths for the GV Tau N
absorption, as assumed in the earlier emission line analysis. 
The column densities of \water, HCN, and \acetylene\ 
in the HV absorption component of GV Tau N 
are 40 times greater on average than 
the emission column densities of AS205 N and DR Tau, and 
the (HCN and \acetylene) column densities of the 
LV component of GV Tau N are $\sim 10$ times greater. 
The roughly order-of-magnitude larger line-of-sight
column densities observed in absorption vs.\ emission 
are expected if the warm molecular layer responsible for 
the T Tauri star emission is viewed at high inclination 
in GV Tau N, as in an edge-on disk.

There is little definitive information available on the orientation
of the GV Tau N disk.  Modeling of MIR interferometric observations
made with VLTI/MIDI favored a high inclination for GV Tau N ($i \sim 80$
degrees; Roccatagliata et al 2011). In addition, the CO overtone
emission from the inner disk of GV Tau N is broad ($\sim 95\,\kms$)
suggesting the disk is not close to face on (Doppmann et al.\ 2008).
In contrast, modeling that fits the spectral energy distribution 
(SED) and scattered light images
of GV Tau N with a disk + envelope model favors a lower inclination
($i \sim 30$ degrees; Sheehan \& Eisner 2014). However, the SED 
modeling may not provide a strong constraint on the orientation of
the inner disk of GV Tau N; because the modeling weights large-scale
phenomena prominently in the fit, the preferred solution may be
influenced primarily by the envelope properties rather than the
inner disk ($< 10$ au) properties where the warm absorption arises.

{\it Clues from Line Profiles.}\quad
While the results above echo previous results from the literature, the high
velocity resolution of the TEXES data lends new insights into the
dynamics of the absorbing gas. The $R$=600 {\it Spitzer} spectra detected
molecular bands but did not resolve individual lines. The NIRSPEC/Keck
spectra studied by Doppmann et al.\ (2008) and Gibb et al.\ (2007)
marginally resolved individual HCN lines in the $L$-band. In comparison,
the TEXES data, with a resolution of $\sim 3\,\kms,$ 
measure the profiles of individual lines.

In interpreting the line profiles, we 
adopt as the system velocity for GV Tau N 
the velocity measured for 
the gaseous envelope surrounding GV Tau 
($v_{\rm helio} = 17.3\pm 0.5 \kms$
or $v_{\rm LSR} = 7.0\,\kms$; Hogerheijde et al.\ 1998).
The stellar radial velocity of GV Tau N is not well known. 
One estimate, based on an absorption component detected in 
the CO overtone emission from the source 
($v_{\rm helio} = 5.8\pm 4.0 \kms$), is blueshifted 
relative to both the cloud and envelope velocity, suggesting 
that the central source may be a spectroscopic binary  
(Doppmann et al.\ 2008).

As described in Section 3, the GV Tau N line profiles have single-dipped, 
$\sim 10 \kms$ FWHM absorption line profiles;   
the absorption core has a velocity offset redward of the system 
velocity and a red wing 
extending to higher velocities. 
The redward offset of the core and red wing 
are not expected for a simple rotating disk and 
suggests gas that is radially inflowing.
The redshifted absorption we observe is reminiscent of earlier 
Keck/NIRSPEC results for GV Tau N. 
Doppmann et al.\ (1998) found that the NIR HCN absorption toward 
GV Tau N, centered at $19 \kms,$ 
is $\sim 2 \kms$ redward of the system
velocity. 

Assuming that all of the absorbing gas resides at $\sim 19 \kms$,
Doppmann et al.\ (2008) argued that the NIR warm molecular absorption
is unlikely to arise from gas in an infalling envelope. Given the 
inferred mass of
the stellar component(s) of GV Tau N ($0.8 \msun$), the envelope
infall velocity $v_{in}=(2G\mstar/r)^{1/2}$ would reach the observed
$2 \kms$ velocity shift at a distance of 360\,au. In contrast, models
of infalling protostellar envelopes predict that the gas temperature
reaches 500\,K only within 2\,au of the star at the $\sim 7\,\lsun$ 
accretion
luminosity of GV Tau (Ceccarelli et al.\ 1996, their Fig.\ 4), i.e.,
well within the distance of 360\,au that would be inferred for the
infalling gas based on its velocity relative to the molecular
envelope.
If the NIR molecular absorption arises in infalling gas at 360\,au, an 
additional, non-traditional source of heating is need to explain 
the observed temperature of the absorbing gas.

A similar argument applies to the MIR absorption. 
At the higher resolution and signal-to-noise of the TEXES spectrum 
compared to the NIRSPEC results, 
we find that the core of the MIR molecular absorption, centered
at 21--22\,\kms, is $\sim 4 \kms$ redward of the system velocity, 
with the red wing of the absorption extending to larger velocities, 
$\sim 15-20 \kms$ redward of the system velocity.
Infalling gas would reach the $\sim 4 \kms$ redshifted velocity 
of the absorption core at
$\sim 90$\,au and the $\sim 17 \kms$ wing velocity at $\sim 5$\,au
from an $0.8 \msun$ star.  While some of the high velocity gas
(arising within $\sim 10$ au) might be warm enough to explain the
observed absorption, the bulk of the infalling gas at $\sim 90$\,au
would be too cool to explain the observed absorption.

Moreover, because of angular momentum conservation, it may be very
difficult for infall to reach the inner few au of a Class I system.
For rotating infall, angular momentum prevents direct accretion
onto the star and its vicinity, instead causing the infalling
material to reach the disk near the centrifugal radius (20--40\,au).
The difficulty of reaching the inner few au through infall is even
more true for late accretion sources in which the infall is
increasingly dominated by high angular momentum material. GV Tau,
which has a weak molecular envelope, is such a late accretion source,
i.e., transitioning from a Class I source to a Class II source.

Another general argument against infall (and in favor of
an inner disk atmosphere) is the molecular composition of the
absorption and its rarity. Figure 15 shows that the GV Tau ratios
of \acetylene, HCN, and water are similar to those of T Tauri  
disk atmospheres seen in emission. Such absorption is rare. 
GV Tau N is the only Class I source known to show \acetylene, HCN, water, 
and \ammonia\ in absorption. One other source 
shows \acetylene\ and HCN absorption in its {\it Spitzer} spectrum 
(IRS46; Lahuis et al.\ 2006). 

If the temperature and composition we
observe in GV Tau N were typical of infalling envelopes, we would
commonly observe these features in Class I sources. 
Compared to a disk atmosphere, gas in an infalling envelope would
cover a much larger fraction of the star, approximately the range of
polar angles between the disk surface and the inner cavity carved
out by outflow. As a result, absorption from infall would be detected
over a much wider range of viewing angles than disks viewed nearly
edge on. The rarity of molecular absorption like that seen in GV Tau N 
argues against its origin in an infalling envelope.

{\it Disk Origin.}\quad 
Unlike the infalling envelope scenario, 
a nearly edge-on rotating disk can account for almost all of the 
properties of the observed molecular absorption: the temperature 
and column density (as discussed above) and its line width.  
Single-dipped absorption line profiles 
can arise when a disk atmosphere is seen in 
absorption against the MIR continuum arising from smaller radii.
While both the line absorption and the continuum emission likely 
arise over a range of radii, for simplicity in the discussion below, 
we assume that the continuum arises from a compact region $r < r_c,$ 
and the absorption occurs at larger radii $r_a > r_c$.
Because the MIR continuum is compact compared to $r_a$, only the
portion of the disk atmosphere at $r_a$ that is close to the
line-of-sight to the continuum is seen in absorption, 
i.e., within an azimuthal angle $\pm \phi$ where $\sin\phi=r_c/r_a$
(Fig.~16)

\begin{figure}[htb!]
  \centering
    \includegraphics[width=0.5\linewidth,trim={1.5in 1.5in 1.0in 2.5in},clip]{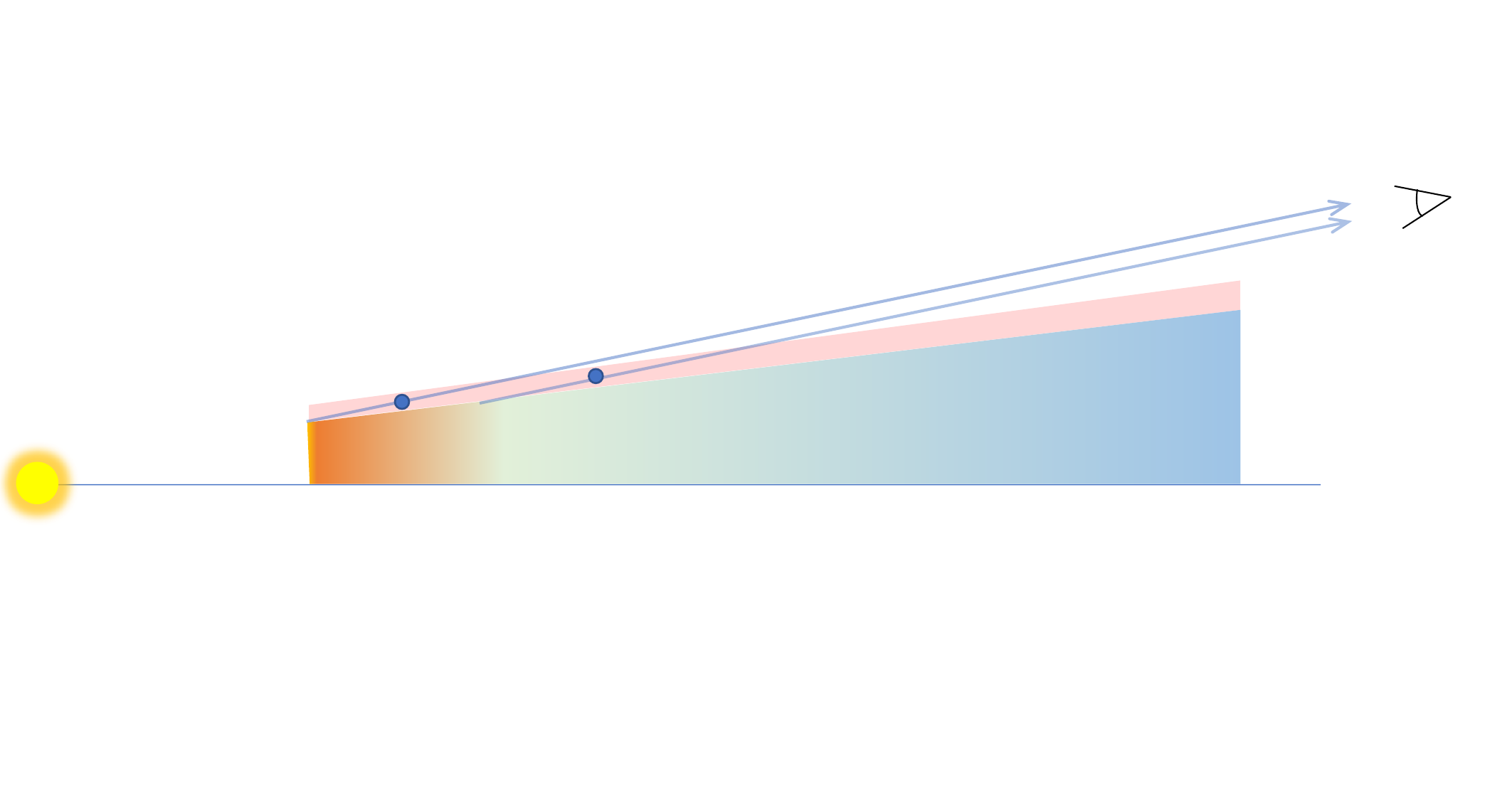}
    \includegraphics[width=0.4\linewidth,trim={2.7in 1.8in 1.0in 2.0in},clip]{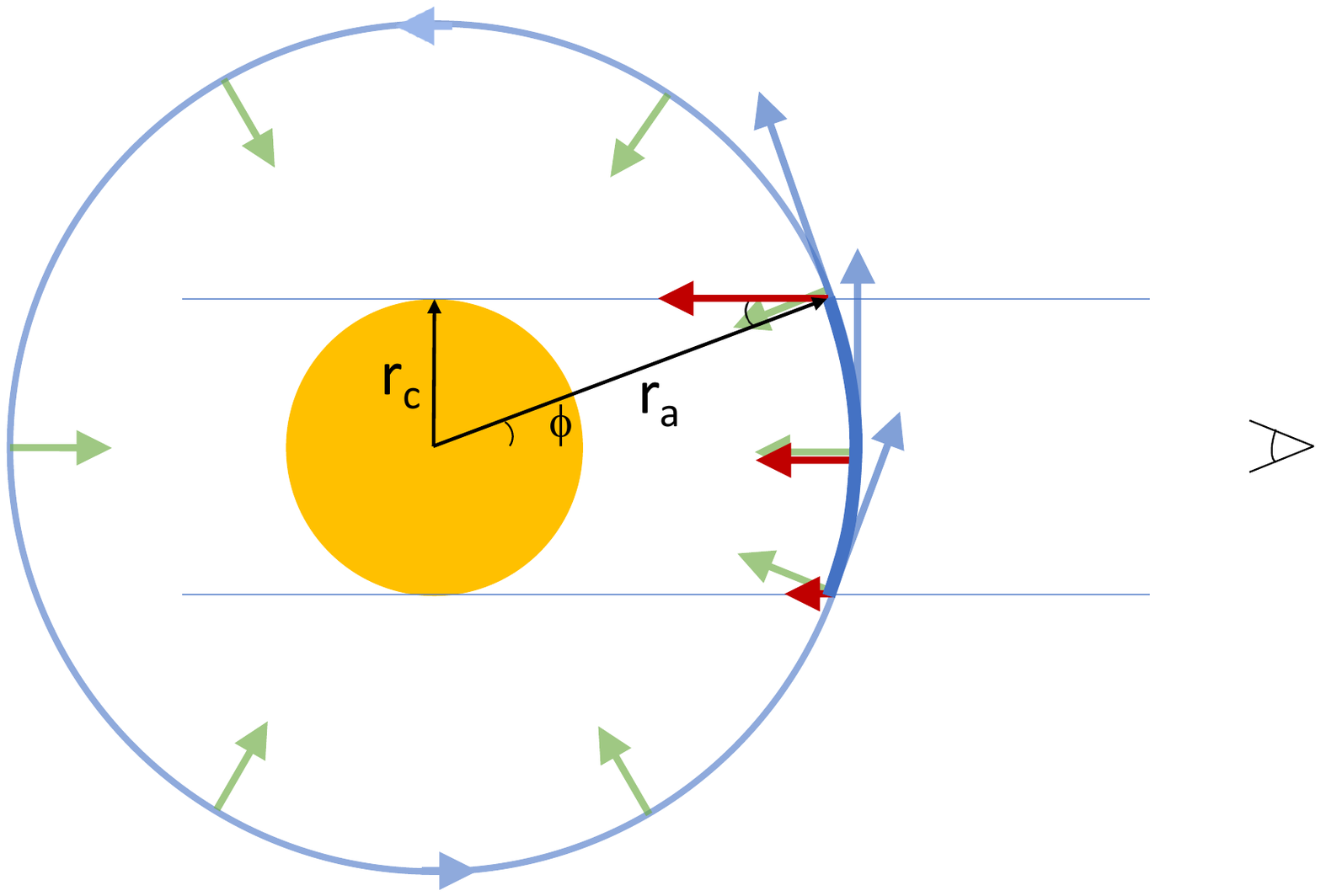}
  \caption{\scriptsize {\it Left:} Schematic (not to scale) 
illustrating the likely observing geometry of GV Tau N. 
The line of sight to the MIR disk continuum (orange region) 
passes through the warm molecular atmosphere at 
larger radii (pink region), producing  
molecular absorption that samples a range of radii 
in the atmosphere (between the blue dots). 
{\it Right:} Top down view of the warm molecular absorption
velocities.  
Absorption by gas at a given radius $r_a$ (heavy blue arc) 
along the line of sight to the warmer MIR continuum 
at $r < r_c$ (yellow-orange region)
has velocity components from both rotation (blue arrows) and 
inflow (green arrows). 
The absorption velocities (red arrows) are shifted to the red 
and span a range of velocities.
}
  \label{fig:fig16}
\end{figure}

Gas at $r_a$ along the line-of-sight to the MIR continuum has a 
maximum projected radial velocity of 
$v_{los} = \pm v_a \sin\phi\,\sin i$ 
where $v_a$ is the disk rotational velocity at $r_a$ and 
$i$ is the disk inclination. 
For a disk in Keplerian rotation, where the projected radial velocity 
due to rotation at $r_c$ is $v_c$ and the projected radial 
velocity at $r_a$ is 
$v_a = v_c \left(r_c/r_a\right)^{1/2},$
$v_{los} = \pm\ v_c \sin i (r_c/r_a)^{3/2}.$
If the 
12\,\micron\ continuum arises within $r_c =0.3$\,au of the star, 
where $v_c = 49\kms$ for an $0.8\msun$ star, 
maximum absorption velocities of 
$v_{los} = \pm 5\kms$ would arise at 
a modestly larger radius of $r_a = 1.4$\,au\,$(\sin i)^{2/3}.$ 
(That is, gas at this radius would absorb background continuum emission 
from $\sim 0.3$\,au over the $10\kms$ FWHM of the observed absorption 
features.) 
Higher velocity absorption would arise from gas at smaller radii. 
Lower velocity absorption arises from both gas at larger radii 
as well as gas at smaller radii that are close to the line of sight 
to the star. 
At a disk radius of 4\,au, the maximum absorption velocity would 
be $\pm 1\kms\,(\sin i).$   

Thus, a conventional rotating, edge-on disk can account for almost 
all of the properties of the observed molecular absorption (the temperature, 
column density, and its line width), except for its redshift from 
the system velocity.  
We can also account for the redward offset of the absorption core  
if the gas in the disk atmosphere is also radially inflowing along 
the disk surface (Fig.~16).  
Such ``surface accretion'' flows are found in simulations of 
rotating magnetized disks. 

\subsection{Surface accretion in GV Tau N?}

In their global ideal MHD simulations of thin disks
threaded by a vertical magnetic field, Zhu \& Stone (2018) found
that accretion occurs primarily in a surface accretion flow in 
the upper, magnetically dominated
disk surface. (See also earlier theoretical work by Stone \& Norman
1994; Beckwith et al.\ 2009; Guilet \& Ogilvie 2012, 2013.) 
The accretion is supersonic, reaching inward velocities 
$v_r = 2$--4\,$c_s$ where $c_s$ is the sound speed.  

In their simulation, 
the rapid accretion at the disk surface (primarily due to the
magnetic $B_r B_\phi$ stress) carries the magnetic field inward, 
pinching it inward at the disk surface. As a result, the disk
atmosphere is connected magnetically to the midplane at larger radii, 
which spins down the atmosphere, producing significantly sub-Keplerian 
rotation---as small as 60\% of Keplerian rotation---in the atmosphere. 
The inward pinching of the magnetic field also launches a disk wind,
although the torque due to the wind at the disk surface plays only
a small role in driving accretion.

The expected properties of the accreting atmosphere are roughly 
consistent with the observed 
properties of the molecular absorption in GV Tau N. 
For molecular gas at 500\,K, supersonic accretion at 
$v_r = 2$--4\,$c_s$ corresponds to $v_r \simeq 3$--5 km/s,  
qualitatively similar to the redshift of the absorption
core ($\sim 4$ km/s); the velocities of the red wing ($\sim 15$--20
km/s) correspond to higher velocities $\sim 13 c_s$.
For gas in sub-Keplerian rotation at a disk radius $r_a$ along the
line of sight to the MIR continuum (which arises within $r_c$), 
its projected radial velocity spans the range 
$v_{los} = \pm\ \beta\,\sin i\ v_c (r_c/r_a)^{3/2}$, as
described above, 
where $\beta$ is a factor that accounts for possible  
deviations from Keplerian rotation. 
If $\beta=0.6$ and $r_c = 0.3$\,au so that 
$v_c=49 \kms$ for a stellar mass of $0.8\msun$, 
an observed absorption core with a FWHM of 
$\pm 5 \kms$ would correspond to absorbing gas at 
$r_a = 1$\,au\,$(\sin i)^{2/3}.$ 

Preliminary work on a detailed radiative transfer model 
of a rotating disk with radial inflow, viewed nearly edge on, 
appears to be able to account for the properties 
of the molecular absorption observed in the GV Tau N spectrum. 
The modeling, which builds on the previous results of Lacy (2013), 
can plausibly account for the shape of the absorption components 
as well as the slightly blueshifted emission seen in the 
\ammonia\ P-branch profiles.  
In one promising scenario, rapidly inflowing gas near the inner rim 
of the disk is seen against the continuum from the inner
rim, and more slowly inflowing gas located near the disk surface 
at larger radii is seen against the disk continuum 
at smaller radii (Lacy et al., in preparation). 

If the observed absorption arises in a surface accretion flow, 
the flow may carry a significant mass accretion rate. 
For an inward flow with velocity $v_{\rm in}$ 
and vertical column density $N_\perp$ 
at a characteristic radius $r_a$, the mass accretion rate is 
\begin{equation}
\dot M_{\rm abs} = 2\pi r_a m_{\rm H}\,v_{\rm in}\,N_\perp.
\end{equation}
If we assume 
\begin{equation}
N_\perp = N_{\rm abs}\eta/x_{\rm mol}
\end{equation}
where $N_{\rm abs}$ is the observed line-of-sight 
molecular absorption column 
(e.g., values in the final column in Tables 5 and 6), 
$\eta$ is the ratio of the vertical to the line-of-sight molecular 
column densities, and $x_{\rm mol}$ is the abundance of the 
tracer molecule relative to hydrogen, 
the associated accretion rate is 
\begin{equation}
\dot M_{\rm abs} = 10^{-9}\msunpery\,
\left({v_{\rm in}\over {4\kms}}\right)
\left({N_{\rm abs} \over 10^{16}\psqcm}\right)
\left({r_a \over 1\,{\rm au}}\right)
\left({\eta \over 0.1}\right)
\left({x_{\rm mol} \over 10^{-6}}\right)^{-1}.
\end{equation}

For the observed HCN absorption, 
the properties of its LV component are 
$v_{\rm in}= 4\kms$, 
$N_{\rm abs} = 7\times 10^{16}\,\psqcm.$
Assuming an abundance in the warm disk atmosphere of 
$ x_{\rm HCN} \sim 10^{-6}$ at $\sim 1$\,au, as found in  
disk chemistry models that account for the MIR molecular emission 
properties of T Tauri disks (Najita \& \'Ad\'amkovics 2017), 
and $\eta = 0.1$, 
the corresponding vertical column density and accretion rate are 
$N_\perp=7\times 10^{21}\,\psqcm$ and 
$\dot M_{\rm abs}\sim 7\times 10^{-9}\msunpery.$

Similarly, given the inferred properties of the 
HV HCN absorption component,  
$v_{\rm in}= 20\kms$, 
$N_{\rm abs} = 3\times 10^{17}\,\psqcm,$
the corresponding vertical column density is 
$N_\perp=3\times 10^{22}\,\psqcm,$ and 
the accretion rate associated with the HV component is 
20 times larger than that of the LV, 
or $\dot M_{\rm abs}\sim 1.5\times 10^{-7}\msunpery.$
Using the properties of the HV water absorption 
($N_{\rm abs} = 5.8\times 10^{19} \psqcm$) and 
a conservative water abundance estimate of 
$x_{\water} \sim 3\times 10^{-4}$ (Najita \& \'Ad\'amkovics 2017) 
yields a similarly large estimate of the accretion rate 
of the HV component, $1 \times 10^{-7}\msunpery.$

In adopting abundances for this estimate, we want to select
values from model atmospheres that are able to match the 
properties of the {\it Spitzer} molecular emission 
from inner disks and use abundance values 
appropriate for the warm atmosphere region that is responsible
for the emission. The models in Najita \& \'Ad\'amkovics (2017) 
do a fair job reproducing the properties of many of the 
molecules detected with {\it Spitzer}. 
To make a conservative estimate of the accretion rate, we have 
adopted model abundances at the high end of the predicted values 
within the layer responsible for the emission. 
The properties of the Walsh et al.\ (2015) and 
Woitke et al.\ (2018) models differ from those of 
Najita \& \'Ad\'amkovics in detail 
(e.g., the molecular atmosphere of Woitke et al.\ is much cooler), 
but their model abundances, if adopted, would imply similar 
or larger accretion rates. 
For example, Walsh et al.\ find a peak HCN abundance of 
$\sim 10^{-7}$ which would imply a much larger accretion 
rate for the LV component of 
$\dot M_{\rm abs}\sim 7\times 10^{-8}\msunpery.$
The peak water abundances for all of the models are 
similar $\sim 10^{-4}.$ 
Thus, the molecular abundances of current disk atmosphere 
models suggest 
large accretion rates for both the LV and HV components. 

The assumed value of $\eta = 0.1$ is plausible for 
the geometry of a highly inclined disk viewed close to edge 
on. A more accurate value could be obtained from a more 
complete model of the physical, thermal, and chemical 
disk conditions. 

The above discussion assumed a local line broadening of 
$2\,\kms$ (Tables 6 and 7) 
and that the full 
absorption width we measure is actually the result of observing 
absorption along multiple lines of sight through the 
rotating disk atmosphere,  
encountering gas at different projected velocities. 
The column densities and accretion rates above would be 
smaller by a factor of two, or larger by a factor of 4, 
if we assumed that the local line broadening is instead
thermal width or the full observed width of 
the absorption component, respectively.

The resulting vertical column densities 
($7\times 10^{21}\,\psqcm$ for LV; 
$2\times 10^{22}\,\psqcm$ for HV) 
are similar to and larger than the 
vertical column densities thought to be responsible for 
the emission from T Tauri disks. 
The resulting accretion rates 
($7\times 10^{-9}\,\msunpery$ for LV; 
$1\times 10^{-7}\,\msunpery$ for HV) 
span the range observed from typical to very active T Tauri stars. 
While there are substantial uncertainties associated with the above
estimates (including the fact that we only measure gas 
columns where the molecular tracers are abundant and atomic 
gas is not probed), the inferred accretion rates
are substantial and fall within the expectations for T Tauri stars.
Thus, the observed radial inflow does appear capable of explaining
the observed accretion rates of very active T Tauri stars and Class
I objects.

The idea of accretion at the disk surface dates back at least
to Gammie (1996), who described a ``layered accretion'' picture of
T Tauri disks, in which only the surface region of the disk that
is sufficiently ionized to couple to magnetic fields participates
in accretion via the magnetorotational instability (the ``active
layer''), and the deeper layers of the disk are a non-accreting
``dead zone.'' In Gammie (1996), the active layer was $\sim
100\,\gcms$ thick, and an equivalent viscosity parameter $\alpha=0.01$
was sufficient to account for observed T Tauri stellar accretion 
rates ($\sim 10^{-8}\,\msunpery$).
In the intervening years, more detailed models of disk ionization 
and our improved understanding of the role of non-ideal MHD 
processes have shrunk theoretical expectations for the vertical 
extent of 
the accreting layer to smaller and smaller column densities, 
requiring larger equivalent values of $\alpha$ 
and larger inflow velocities to deliver the same accretion rate.
Here T Tauri-like accretion rates appear to be transported over 
vertical columns of only $\sim 0.01\,\gcms.$ 

Supersonic surface accretion flows---which differ 
from previous disk accretion mechanisms in that 
accretion is driven not by turbulence or a 
wind, but primarily by large-scale net magnetic fields 
in the disk atmosphere (Zhu \& Stone 2018)---may 
help resolve the open question of how 
protoplanetary disks accrete at planet formation distances 
($\sim$0.3--10\,au). 
Because of the low ionization fractions of disks  
below their surfaces, it is increasing unclear whether  
turbulence generated by the magnetorotational instability 
(MRI) can drive a significant accretion rate in 
protoplanetary disks 
(e.g., Turner et al.\ 2014). 

The apparent evidence for a supersonic surface accretion flow in 
the disk of GV Tau N suggests an alternative pathway for disk 
accretion. 
As a mechanism that redistributes angular momentum within the 
disk rather than removing it from the 
system---as in a disk wind---the supersonic surface 
accretion flows studied by Zhu \& Stone (2018) 
contribute to disk spreading: angular momentum 
from the disk surface is transported to the midplane at larger 
radii, which induces the surface to accrete and the midplane 
to spread to larger radii. As an ``in-disk'' transport mechanism, 
such surface accretion flows may help explain the larger sizes 
of gas disks surrounding Class II sources 
(as large as $\sim 500-800$\,au) 
compared to those of Class I sources (typically $< 100$\,au). 
The striking size difference has been interpreted 
as evidence that  some ``in disk'' angular momentum transport 
process is active in the Class II (T Tauri) phase 
(Najita \& Bergin 2018).

Our results may also be related to magnetothermal 
wind-driven accretion. 
Because the winds rely on relatively weak magnetic fields, 
their magnetic lever arms are small; as a result, 
transporting disk material from large radii to small 
(over a factor of 30 in radius, from 10\,au to 0.3\,au) 
is an inefficient process requiring large wind mass loss rates,  
comparable to or greater than the accretion rate (Bai \& Stone 2013).  
Although strong observational evidence for such massive, 
angular momentum-removing disk winds is currently lacking, 
the winds are also predicted to produce accretion 
in narrow current sheets, which can reach high velocities 
near the disk surface under certain conditions (Bai 2013). If 
the net vertical field is anti-aligned with disk rotation,  
near-sonic to supersonic inflow can develop on one side 
of the disk surface 
over some range of radii; under other conditions, the 
accretion flow reaches much lower velocities or occurs 
close to the midplane (Bai \& Stone 2013, Bai 2017). 
The accretion flow we observe in GV Tau N may be related 
to the predicted high-velocity flows.  

Thus, our spectroscopic results 
appear to capture disk accretion in action 
and provide observational support for supersonic surface accretion, 
a potentially important mode of accretion in protoplanetary disks.
To explore this possibility further, 
we need future spectroscopic searches for redshifted warm molecular
absorption from other Class I sources; the incidence rate of the 
absorption, which constrains its covering fraction,  
can distinguish between an origin in a disk atmosphere 
(small covering fraction) or an infalling envelope 
(large covering fraction).
Detailed radiative transfer (RT) modeling is needed to understand 
whether the line profiles we observe can actually be produced 
in disk atmospheres undergoing surface accretion. 
Similarly, thermal-chemical-RT modeling of infalling envelopes 
is needed to explore that alternative scenario as an explanation 
for the observed line profiles.  

In addition, future theoretical work is needed to understand whether 
surface accretion flows can be driven under realistic ionization 
conditions in protoplanetary disks; the Zhu \& Stone (2018) 
simulations were carried out assuming ideal MHD.  
At the observed inflow velocities of $\sim 5\,\kms$, the 
accreting material will travel an au in a year, implying that 
the flow must be rapidly replenished in order to sustain it 
over protostellar lifetimes ($\sim 10^5$ yr). 
Whether and how the replenishment 
occurs---from disk inflows at larger radii and/or from deeper 
in the disk---is an open question. 
The results reported by Fuente et al.\ (2020) may provide 
a clue. They find evidence for 
redshifted absorption at modest inflow velocities 
($\lesssim 3\,\kms$) in cool molecular gas at larger 
distances from GV Tau N, as traced in $^{13}$CO (J=3-2).

It would also be useful to obtain additional observational 
evidence for surface accretion flows. 
As discussed above, even 
if disks do commonly accrete through their sub-Keplerian surfaces
at supersonic speeds, such flows are unlikely to be commonly
observed. Nearly edge-on disks, like that inferred for GV Tau N,
are advantageous systems in which to search for surface accretion
flows because the modest inward motions (few times the sound speed)
are more readily detected from that viewing angle. 
However, edge-on disks are intrinsically
rare because of their special orientation. Consistent with this
picture, GV Tau N is one of the few YSOs to show molecular absorption
in {\it Spitzer}/IRS spectra, and yet its SED is similar to that of other Class
I sources (Furlan et al.\ 2008). That is, GV Tau N appears to be a 
typical Class I source viewed at an unusual inclination.

Although GV Tau N is rare in its molecular absorption properties,
radial inflows have been reported in several other disk systems.
As described by Zhang et al.\ (2015), 
the classical T Tauri star AA Tau, whose inner disk is highly inclined 
($\sim 70$--75 degrees), 
shows very broad CO fundamental emission lines,
with narrow absorption superposed near the line center. 
Following a photometric dimming event in 2011, 
the molecular absorption component increased in strength
and showed a constant redshift of $\sim 6\,\kms$ with 
respect to the star.
The observed velocity shift, 
temperature of the absorbing gas ($\sim 500$\,K), 
and inferred column density of the absorber 
($N_{\rm H} \sim 3\times 10^{22}\,\psqcm$)
are comparable to the properties of the absorption in GV Tau N.

Perhaps most dramatically, Boogert et al.\ (2002) reported redshifted
absorption in the 5\,\micron\ CO absorption spectrum of the Class I
protostar L1489 IRS. The CO spectrum revealed redshifted absorption
profiles similar to those seen here, but with a red wing extending
to $100\,\kms$  produced by warm ($\sim 250$\,K) gas. The absorption
was interpreted as inward flowing gas at the warm disk surface,
although at the time the phenomenon was reported, the physical 
origin of the gas was unclear. It is tempting to speculate that 
the CO absorption in L1489 IRS also arises in a disk surface 
accretion flow, albeit one with a very high inflow velocity.

As another possible example of a surface accretion flow 
but on a larger scale, 
spatially resolved CO $J$=3--2 imaging with ALMA of the Herbig Ae
star HD100546 shows deviations from Keplerian rotation which have
been interpreted as indicating either a severely warped and twisted
inner disk or radially infalling gas within 100\,au (Walsh et al.\
2017). The lack of evidence for a disk warp from high contrast
imaging of the dust disk favors the latter explanation. 
To explain the CO spatial and spectral structure, 
the required inward velocities are several times the sound speed, 
similar to the inflow speeds found in simulations that report 
surface accretion flows (Zhu \& Stone 2018). 
Walsh et al.\ found reasonable fits to the HD100546 data with 
a radial flow that is 63\% of the Keplerian velocity within 84\,au. 
Given the $\sim 4.6\,\kms$ Keplerian velocity of HD100546 at 84\,au 
(assuming a $2\msun$ star), the radial flow velocity is $2.6\kms.$
If the gas temperature in the disk atmosphere at 84\,au is 30\,K, 
the expected inward flow velocity from a surface accretion flow is 
approximately $4 c_s = 2.2\kms,$ similar to the radial flow velocity
inferred from the observations. 

Similarly, 
ALMA imaging of HCO$^+$ emission from the classical T Tauri star AA
Tau has a twist (within the innermost ring at  40\,au) in the projected
velocity field relative to the velocity field at larger radii, which
is also interpreted as a warp or an inward radial flow (Loomis et al.\ 
2017). Thus, although radial inflows have not been much reported 
to date, studies of detailed disk dynamics with ALMA and high 
resolution studies of nearly edge-on disks, carried out for 
larger samples than have been explored to date, can clarify 
this picture. 

Another way to detect surface accretion flows may be through their
sub-Keplerian rotation. While disk rotation at planet formation
distances ($< 10$\,au) has been demonstrated using velocity-resolved
molecular emission line profiles (e.g., CO fundamental emission), 
demonstrating that the rotation is sub-Keplerian requires spatial
constraints on the observed velocities (i.e., spatially and spectrally
resolved emission or spectroastrometry; 
e.g., Pontoppidan et al.\ 2011)  
as well as {\it independently determined} stellar masses. The latter 
may be challenging to 
obtain. One of the ``gold standard'' methods for measuring 
pre-main-sequence stellar masses is to spatially resolve the 
rotation of outer disks assuming pure Keplerian 
rotation (e.g., Simon et al.\ 2000).

\subsection{First Detection of \ammonia\ in an Inner Disk?}

Although 
ammonia has been previously reported in the outer disk of one young star 
(TW Hya, in data taken with the {\it Herschel Space Observatory}; 
Salinas et al.\ 2016), it has not been previously reported in 
inner disks, in either emission or absorption.
In their analysis of {\it Spitzer} molecular emission spectra taken at low resolution, 
Salyk et al.\ (2011) reported upper limits on the column density of ammonia,
based on simple slab modeling and an assumed temperature of 400\,K. 
The results correspond to upper limits on the ratio of ammonia to water of 
[\ammonia/\water] $\lesssim 0.005$.
The analysis by Bast et al.\ (2013) of the {\it Spitzer} absorption 
spectrum of GV Tau led to an upper limit on its \ammonia\ absorption
column density of $2\times 10^{16}\psqcm$ assuming a temperature of 500\,K. 
Our measured absorption columns for \ammonia\ towards GV Tau N are
consistent with these upper limits.

Ammonia upper limits have also been derived from high resolution
spectra of inner T Tauri disks. 
Mandell et al.\ (2012) reported upper limits on warm \ammonia\ 
emission corresponding to [\ammonia/\water] $\lesssim 0.2,$  
based on VLT/CRIRES spectra at 3\micron.  
Summarizing the results of a Gemini/TEXES program to search for 
\ammonia\ Q-branch emission at 10.75 \micron\ from inner T Tauri disks,
Pontoppidan et al.\ (2019) found a lower average abundance upper limit
of [\ammonia/\water] $< 0.003$.
When compared with the HCN abundance measured in the same disks,  
the \ammonia\ upper limit corresponds to [\ammonia/HCN] $< 0.1.$  

Here we find a much larger [\ammonia/HCN] ratio for GV Tau N, 
which shows comparable absorption column densities in \ammonia\ 
and HCN. Using the best constrained
values, that of $f_d\,N$ for the low velocity component of each 
molecule at each epoch,
our inferred ratio of [\ammonia/HCN] is $\sim 0.5,$
much larger than the [\ammonia/HCN]$\sim 0.1$ upper limits obtained  
by Pontoppidan et al.\ (2019)
from the molecular emission spectrum of three T Tauri disks. 

Whereas conspicuous \ammonia\ emission is absent in the 
{\it Spitzer} IRS spectra of T Tauri stars 
(e.g., Salyk et al.\ 2011, Carr \& Najita 2011), detectable emission 
is expected if \ammonia\ is co-located with the HCN (i.e., 
has the same temperature; $\sim 600$\,K in emission) 
and has an abundance similar to that of HCN.  
To illustrate this discrepancy, we show in Figure 17 examples of 
predicted molecular emission spectra for simple slab models 
of T Tauri disks.
The predictions adopt the HCN column densities 
that have been derived for these disks assuming that the HCN 
emission has the same temperature and emitting area 
as the \water\ emission  
(Carr \& Najita 2011; Najita et al.\ 2018).
At the GV Tau N column density ratio of [\ammonia/HCN]=0.5 
and a temperature of 600\,K,
the resulting \ammonia\ emission would be measureable 
(solid red line).
At the lower ratio of [\ammonia/HCN]=0.1, corresponding to 
the high-resolution limits from Pontoppidan et al.\ (2019), 
the \ammonia\ emission would not be detectable 
(blue line).

\begin{figure}[htb!]
  \centering
    \includegraphics[width=0.5\linewidth,trim={0.0in 0.5in 0.0in 0.5in},clip]{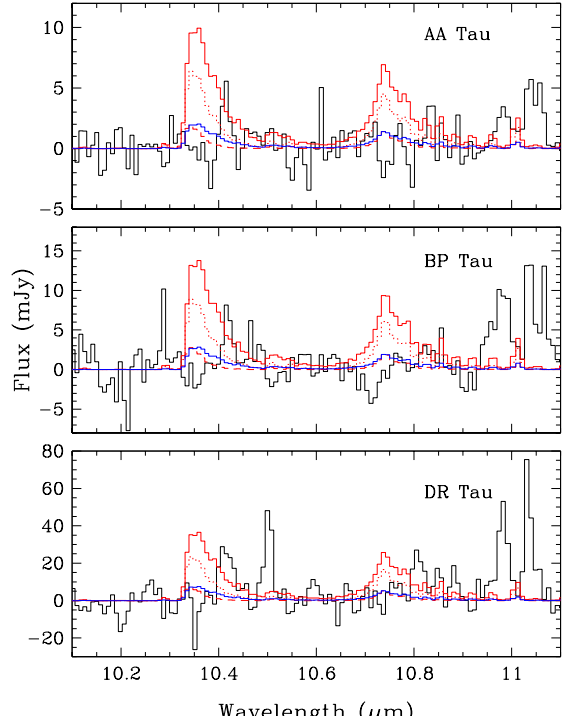}
  \caption{\scriptsize Observed spectra (black line) compared with simple slab emission 
models for \ammonia\ emission from inner T Tauri disks assuming
[\ammonia/HCN] abundance ratios of 0.5 (red lines) and 0.1 (blue line)
and HCN columns from Carr \& Najita (2011) that
assume optically thin emission, i.e., the HCN emission has
the same emitting area as the water emission.
Models at the higher abundance ratio are for temperatures of
600\,K (solid red), 450\,K (dotted red), and 300\,K (dashed red);
the model with the lower abundance ratio assumes 600\,K.
The three stars shown are ordered from low (AA Tau) to medium (BP Tau) to
high (DR Tau) stellar accretion rate.
}
  \label{fig:fig17}
\end{figure}

The much higher [\ammonia/HCN] column density ratio observed in 
absorption GV Tau N compared to that seen in emission in T Tauri 
disks might be expected if the emission and absorption features 
probe the conditions at different disk heights. 
In their thermal-chemical models of irradiated disk atmospheres,
Najita \& \'Ad\'amkovics (2017) find that \ammonia\ is abundant deeper in the
atmosphere, and at lower gas temperature, 
than the region in which HCN and \acetylene\ are abundant.
Thus, we expect a low [\ammonia/HCN] ratio in the HCN-emitting gas 
(Figure 17, blue line). 
Abundant but cool \ammonia\ located below the HCN-emitting 
gas would also produce weak to negligible emission 
(Figure 17, dashed red line). 

In contrast, cool gas could still be readily detected in absorption in transitions
out of the ground vibrational state, 
particularly with the large slant column
densities for a disk viewed at high inclination.
The cooler temperatures we measure for the 
\ammonia, HCN, and \acetylene\ absorption (450\,K), 
compared to the typical temperatures of the 
HCN and \acetylene\ emission from T Tauri stars 
(600-1200\,K; Carr \& Najita 2011; Salyk et al.\ 2011) 
are consistent with the idea that the GV Tau N absorption
probes a deeper layer in the disk atmosphere than the region
responsible for the MIR molecular emission from T Tauri disks.
The low temperatures we find for the \ammonia\ absorption 
(250\,K in 2007 and 450\,K in 2006) are roughly consistent 
with this explanation for the lack of detectable \ammonia\ emission
from T Tauri disks 
and with 
the cooler temperatures anticipated for \ammonia\ compared to HCN from the 
disk thermal-chemical models.

The measured column densities of \ammonia\ and HCN in GV Tau N 
add to our current understanding of the nitrogen reservoir in disks. 
As described by Pontoppidan et al.\ (2019), nitrogen is highly 
depleted in the bulk Earth, by 5--6 orders of magnitude, 
compared to its cosmic abundance. 
The low abundance suggests that the bulk carrier of nitrogen in 
the material that formed the Earth was much more volatile than water, 
favoring a nitrogen-bearing molecule like N$_2$, which has 
a low binding energy (430\,K) compared to other potentially 
abundant molecules such as HCN and \ammonia\ 
(3610\,K and 3130\,K respectively; Walsh et al.\ 2015). 
Our results for GV Tau N are consistent with this picture. 
Because the measured column of \ammonia\ is modest, only comparable to 
that of HCN, it cannot be a major missing reservoir of nitrogen, 
and a molecule like N$_2$ is instead the likely dominant nitrogen 
reservoir.

\section{Summary and Conclusions}

The mid-infrared spectra of GV Tau N reported here were obtained
with the original goal of studying the physical properties and
molecular content of a disk viewed edge-on.  The opportunity to study 
an unusually large column density of disk gas in absorption 
offered the potential to detect new molecular species. Consistent
with that expectation, the TEXES spectra revealed the first
evidence for \ammonia\ in the planet formation region of disks.
The measured temperatures, molecular column densities, and 
column density ratios of the detected species 
(\acetylene, HCN, \ammonia, and \water)  
are consistent with the properties of a disk atmosphere within 
a few au of the star viewed at high inclination.  
While the \ammonia\ abundance measured here is higher than the upper
limits obtained from molecular emission studies of disks, our results
do confirm the expectation that \ammonia\ is not a major missing
reservoir of nitrogen. If, as expected, the dominant nitrogen
reservoir in inner disks is instead N$_2$, its high
volatility would make it difficult to incorporate into forming
planets, a situation that may help to explain the low nitrogen
content of the bulk Earth.

More interestingly, the TEXES spectra reveal an unexpected and
significant redshift to the detected molecular absorption features, 
indicative of inflow at the disk surface.
From the properties of the molecular absorption 
(column density, velocity shift), we can infer that the 
redshifted absorption carries a significant accretion rate:  
$\dot M_{\rm abs} \sim 10^{-8} -10^{-7}\msunpery,$
comparable to the stellar accretion rates of active T Tauri stars. 
Thus we may be observing disk accretion in action. 
The results may provide observational evidence
for a new disk accretion pathway for young protoplanetary disks: 
supersonic ``surface accretion flows.'' These flows have been 
found in MHD simulations of magnetized disks 
(e.g., Zhu \& Stone 2018), but their potential 
role in young protoplanetary disks has received limited attention 
to date. 
The observed flows may also be related to accretion flows 
generated by magnetothermal winds. 
Future spectroscopy of the dynamics of other edge-on disks would 
help establish whether supersonic inward flows are common among 
young disks. 
In addition, 
future simulations are needed to understand whether supersonic 
surface accretion flows can be sustained under realistic ionization 
conditions in protoplanetary disks.

\appendix
\section{TEXES Spectra of GV Tau N}

The figures in this section show the entire set of spectra of GV
Tau N used in this study. Each panel shows the pipeline-reduced
spectra on the observed wavelength scale 
before additional corrections were made to remove low-order
structure in the continuum. Detected lines are annotated at the 
velocity of the gaseous envelope surrounding GV Tau, 
$v_{\rm helio} =17.3\,\kms$ 
(or $v_{\rm LSR} = 7.0\,\kms$; 
Hogerheijde et al.\ 1998). 
The molecular absorption features are clearly redshifted 
with respect to the envelope velocity.

\begin{figure}
  \figurenum{A1a}
  \centering
    \includegraphics[width=0.9\linewidth,trim={0.0in 0.0in 0.0in 0.0in},clip]{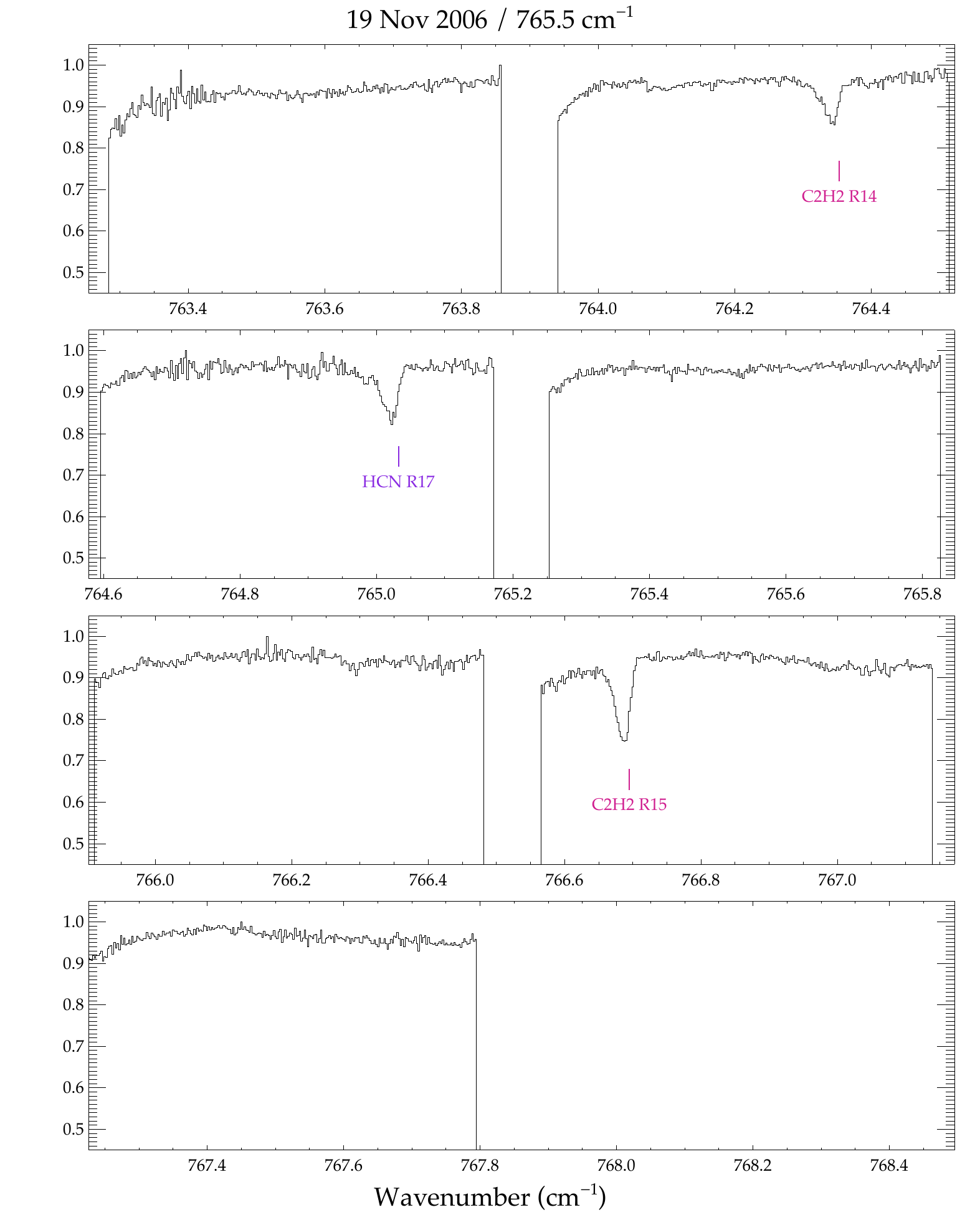}
  \caption{TEXES pipeline-reduced spectra of GV Tau N, shown before additional corrections were made to remove low-order structure in the continuum.  
Absorption lines of \acetylene\ (red), HCN (purple), \ammonia\ (green), and water (blue) are marked in each panel, as is the 12\,\micron\ H$_2$ emission line (blue). 
This panel shows the GV Tau N spectrum  
observed in 2006 at the setting centered at 765.5\,\icm. 
}
  \label{fig:figA1a}
\end{figure}

\begin{figure}
  \figurenum{A1b}
  \centering
    \includegraphics[width=0.9\linewidth,trim={0.0in 0.0in 0.0in 0.0in},clip]{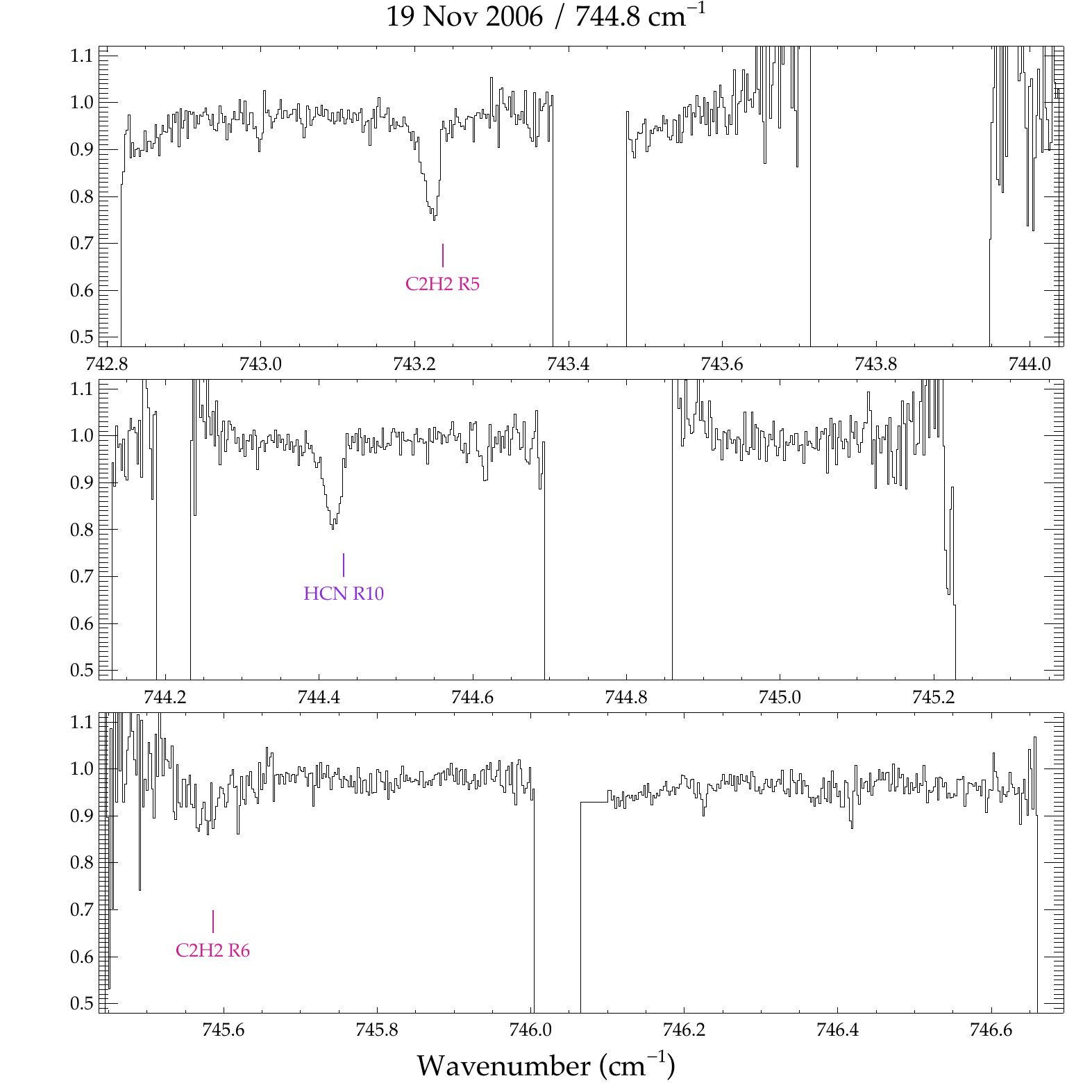}
  \caption{As in Figure A1a, for the setting centered at 744.8\,\icm\ observed in 2006. 
}
  \label{fig:figA1b}
\end{figure}

\begin{figure}
  \figurenum{A1c}
  \centering
    \includegraphics[width=0.9\linewidth,trim={0.0in 0.0in 0.0in 0.0in},clip]{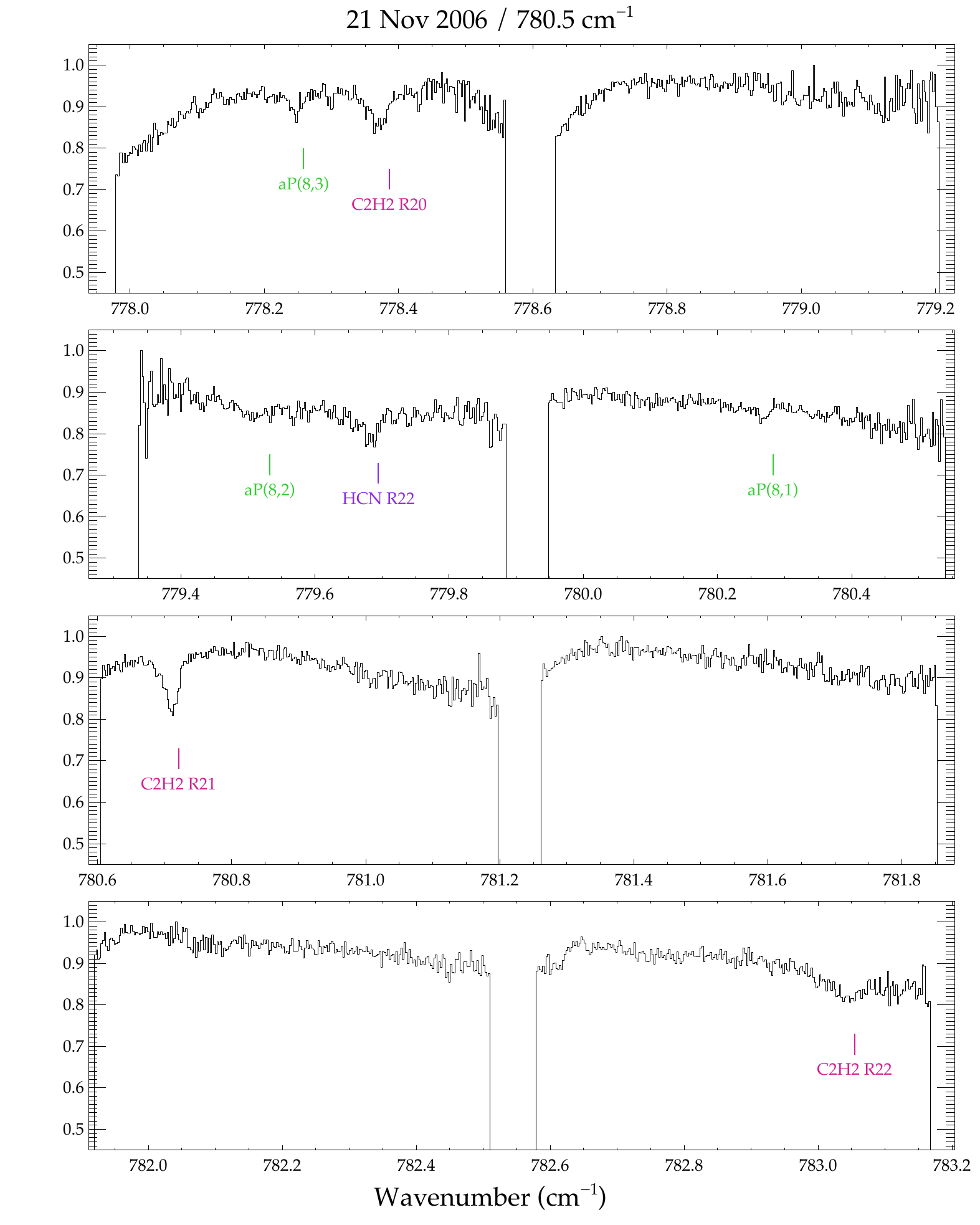}
  \caption{As in Figure A1a, for the setting centered at 780.5\,\icm\ observed in 2006. 
}
  \label{fig:figA1c}
\end{figure}

\begin{figure}
  \figurenum{A1d}
  \centering
    \includegraphics[width=0.9\linewidth,trim={0.0in 0.0in 0.0in 0.0in},clip]{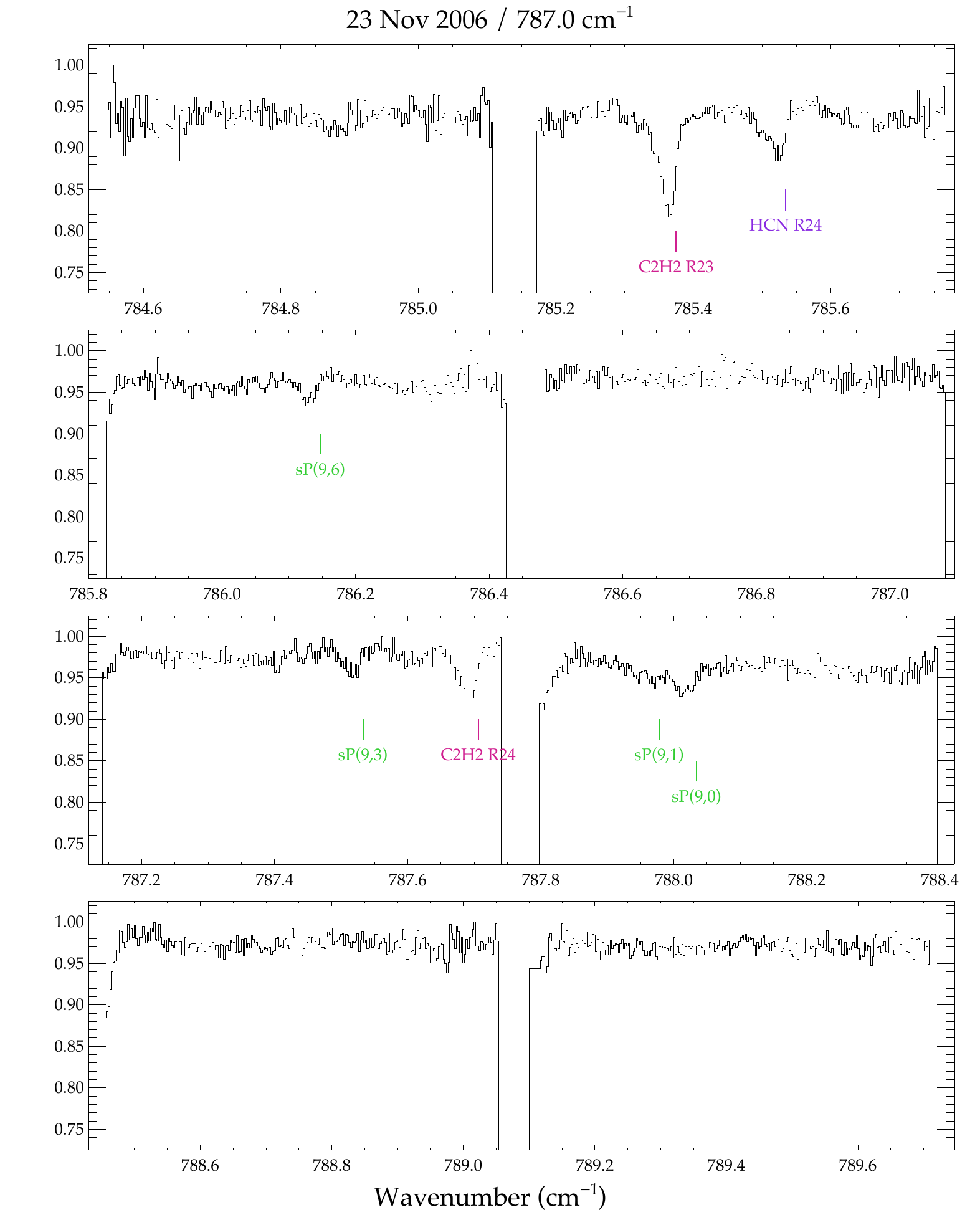}
  \caption{As in Figure A1a, for the setting centered at 787.0\,\icm\ observed in 2006. 
}
  \label{fig:figA1d}
\end{figure}

\begin{figure}
  \figurenum{A1e}
  \centering
    \includegraphics[width=0.9\linewidth,trim={0.0in 0.0in 0.0in 0.0in},clip]{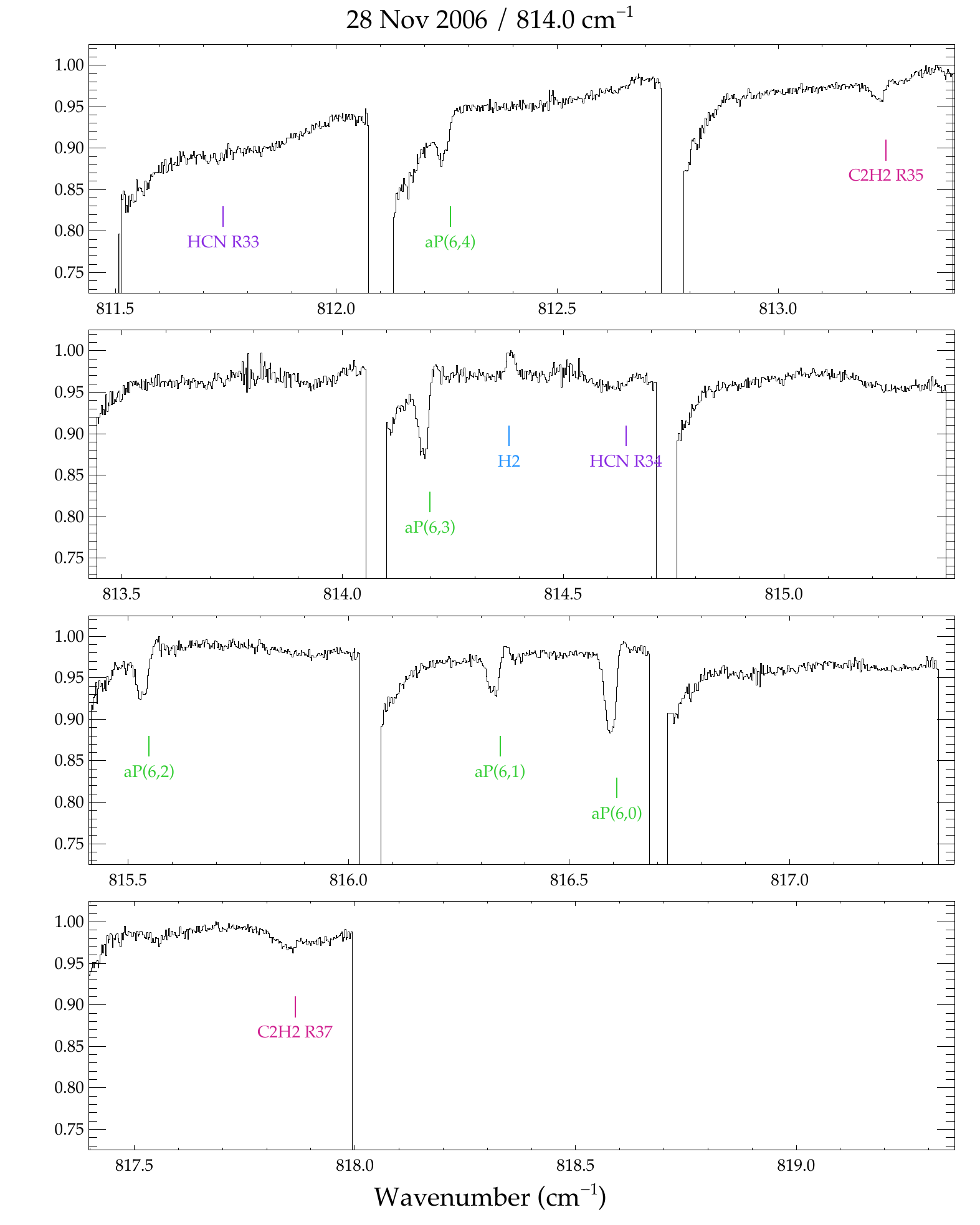}
  \caption{As in Figure A1a, for the setting centered at 814.0\,\icm\ observed in 2006. 
}
  \label{fig:figA1e}
\end{figure}

\begin{figure}
  \figurenum{A1f}
  \centering
    \includegraphics[width=0.9\linewidth,trim={0.0in 0.0in 0.0in 0.0in},clip]{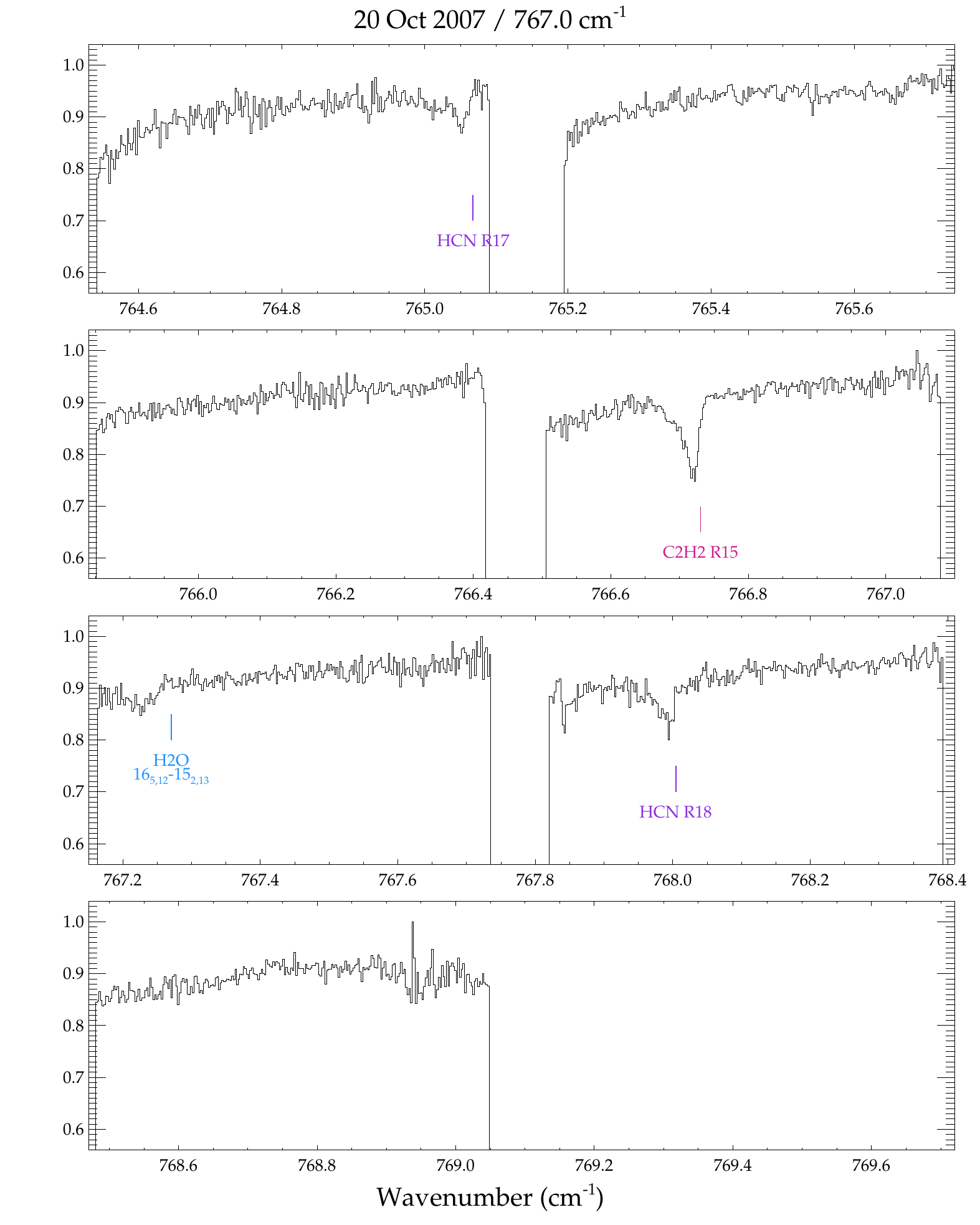}
  \caption{As in Figure A1a, for the setting centered at 767.0\,\icm\ observed in 2007. 
}
  \label{fig:figA1f}
\end{figure}

\begin{figure}
  \figurenum{A1g}
  \centering
    \includegraphics[width=0.9\linewidth,trim={0.0in 0.0in 0.0in 0.0in},clip]{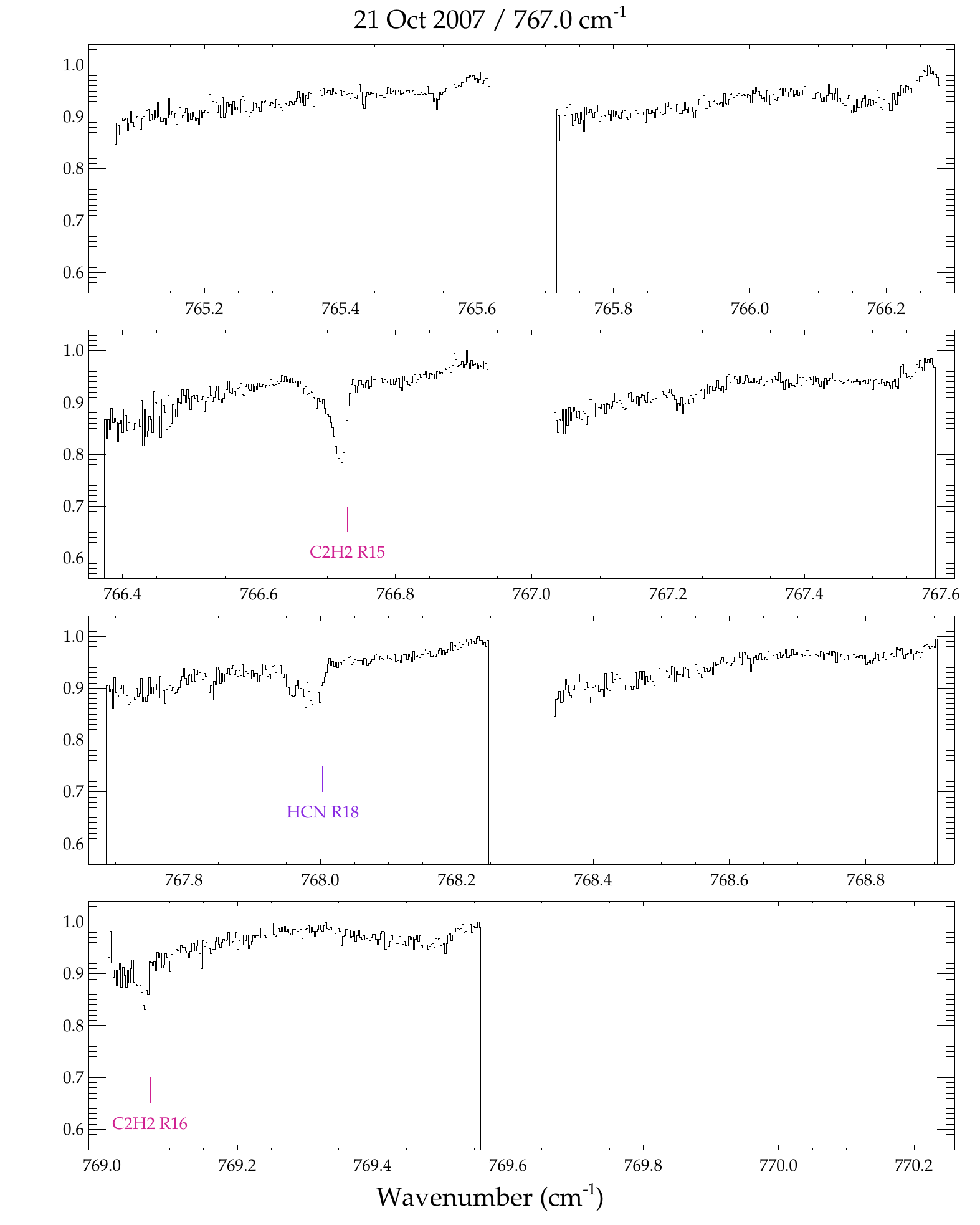}
  \caption{As in Figure A1a, for the setting observed in 2007 
centered one order to the red of the other setting at 767.0\,\icm\ 
shown in the previous panel. 
}
  \label{fig:figA1g}
\end{figure}

\begin{figure}
  \figurenum{A1h}
  \centering
    \includegraphics[width=0.9\linewidth,trim={0.0in 0.0in 0.0in 0.0in},clip]{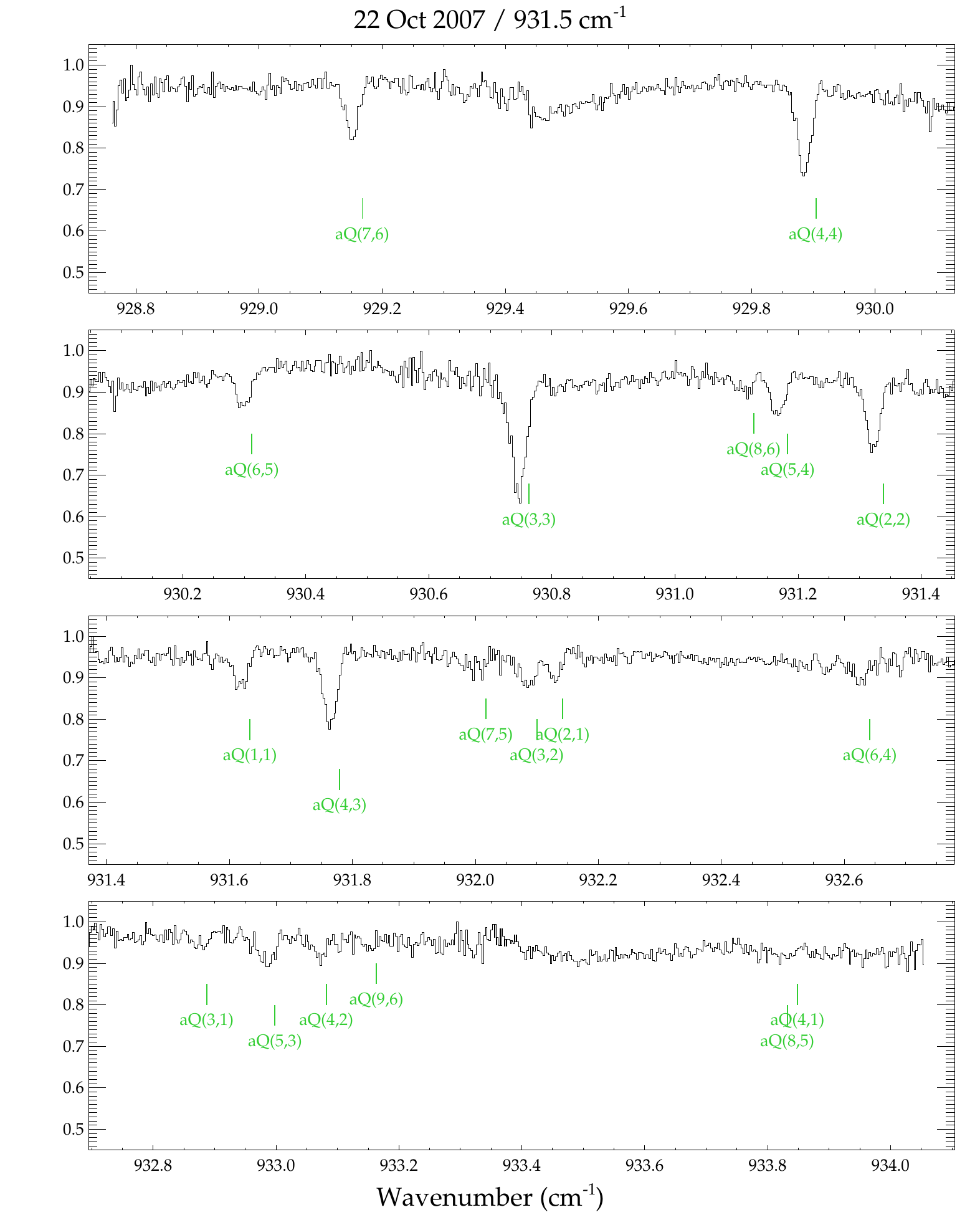}
  \caption{As in Figure A1a, for the setting centered at 932.0\,\icm\ observed in 2007. 
}
  \label{fig:figA1h}
\end{figure}

\begin{figure}
  \figurenum{A1i}
  \centering
    \includegraphics[width=0.9\linewidth,trim={0.0in 0.0in 0.0in 0.0in},clip]{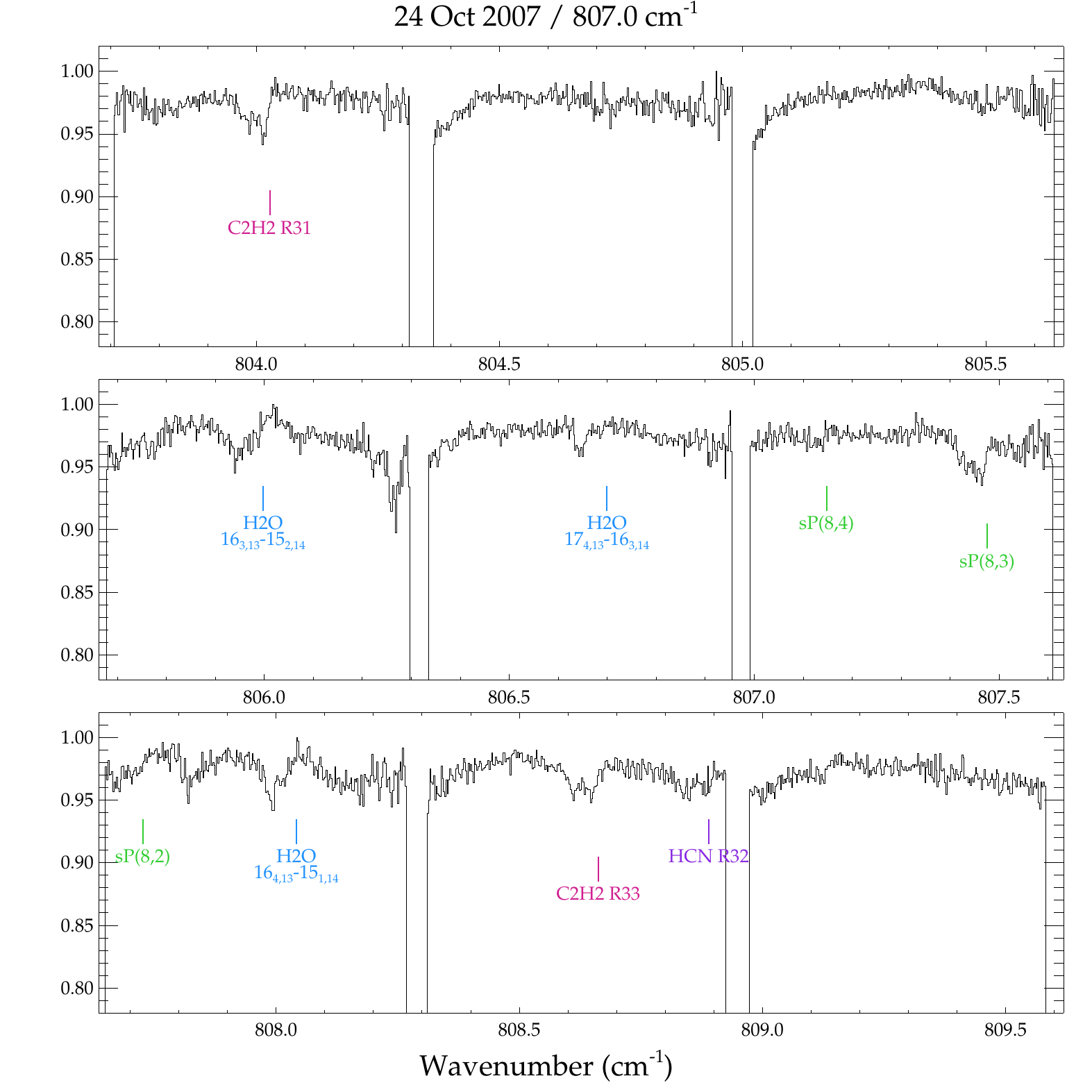}
  \caption{As in Figure A1a, for the setting centered at 807.0\,\icm\ observed in 2007. 
}
  \label{fig:figA1i}
\end{figure}

\begin{figure}
  \figurenum{A1j}
  \centering
    \includegraphics[width=0.9\linewidth,trim={0.0in 0.0in 0.0in 0.0in},clip]{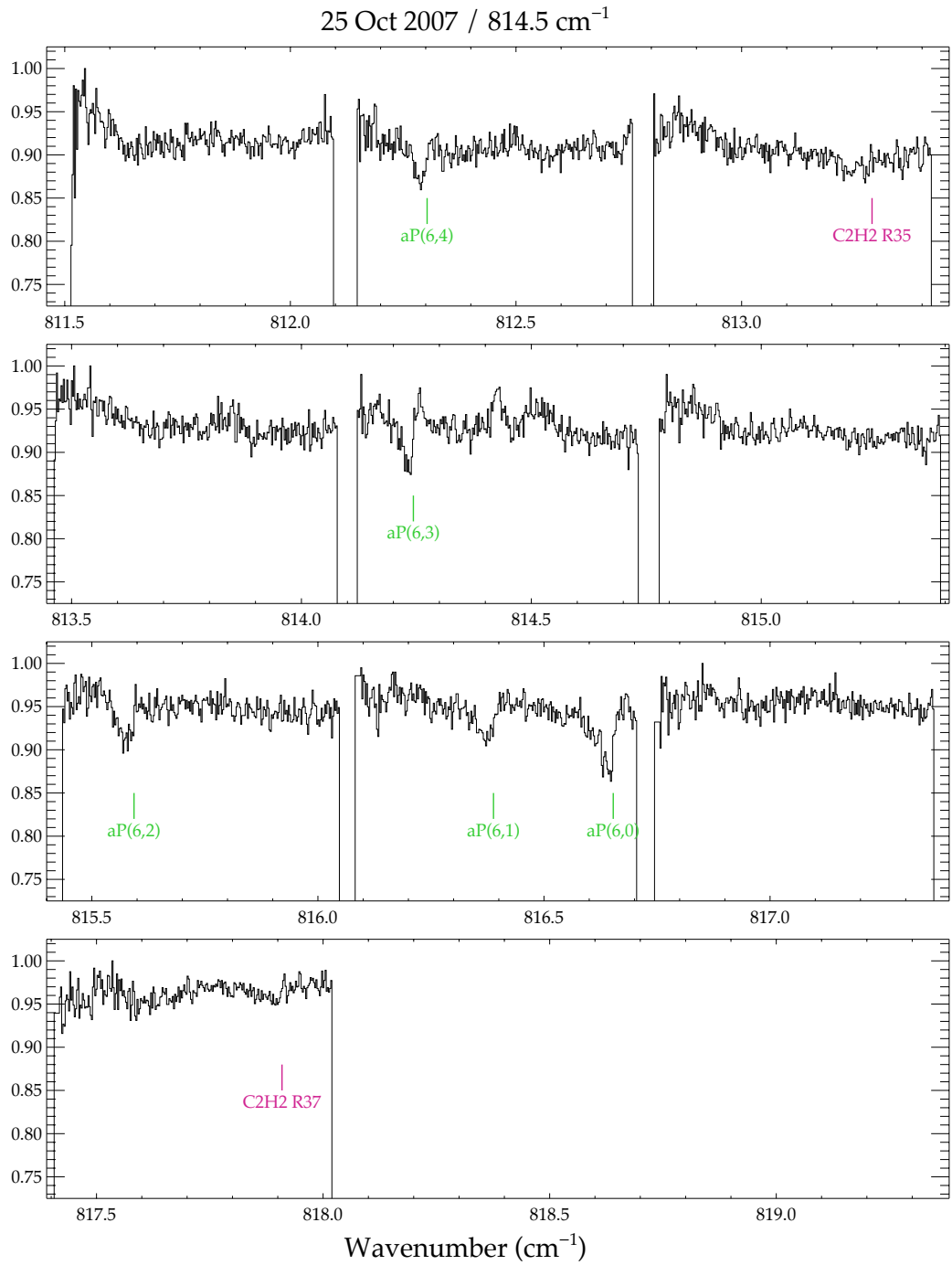}
  \caption{As in Figure A1a, for the setting centered at 814.5\,\icm\ observed in 2007. 
}
  \label{fig:figA1j}
\end{figure}

\newpage

\acknowledgments
We gratefully thank Zhaohuan Zhu for introducing us to his 2018 
paper and for helpful discussion and advice.
We also thank the referee for helpful comments that 
improved the manuscript. 
This work was performed in part at the Aspen Center for Physics which is 
supported by the National Science Foundation grant PHY-1607611. Work by 
SDB was performed in part at the National Optical Astronomy Observatory. 
NOAO is operated by the Association of Universities for Research in Astronomy 
(AURA), Inc. under a cooperative agreement with the National Science Foundation. 
SDB also acknowledges support from this work by NASA Agreement No. NXX15AD94G; 
NASA Agreement No. NNX16AJ81G; and NSF-AST 1517014.

The authors wish to recognize and acknowledge the very significant
cultural role and reverence that the summit of Mauna Kea has always
had within the indigenous Hawaiian community. We are most fortunate
to have the opportunity to conduct observations from this mountain.
We thank the Gemini staff, and John White in particular, for their
support of TEXES observations on Gemini North. The development of
TEXES was supported by grants from the NSF and the NASA/USRA SOFIA
project. Modification of TEXES for use on Gemini was supported by
Gemini Observatory. Observations with TEXES were supported by NSF
grant AST 06-07312. M.\ J.\ R.\ acknowledges support from NSF grant
AST 07-08074, NASA grant NNG04GG92G, and NASA Collaborative Agreement 
80NSSC19K1701.

This work is based on observations obtained at the international
Gemini Observatory, a program of NSF’s NOIRLab, which is managed
by the Association of Universities for Research in Astronomy (AURA)
under a cooperative agreement with the National Science Foundation
on behalf of the Gemini Observatory partnership: the National Science
Foundation (United States), National Research Council (Canada),
Agencia Nacional de Investigaci\'{o}n y Desarrollo (Chile), Ministerio
de Ciencia, Tecnolog\'{i}a e Innovaci\'{o}n (Argentina), Minist\'{e}rio
da Ci\^{e}ncia, Tecnologia, Inova\c{c}\~{o}es e Comunica\c{c}\~{o}es
(Brazil), and Korea Astronomy and Space Science Institute (Republic
of Korea).



\includepdf[pages=-]{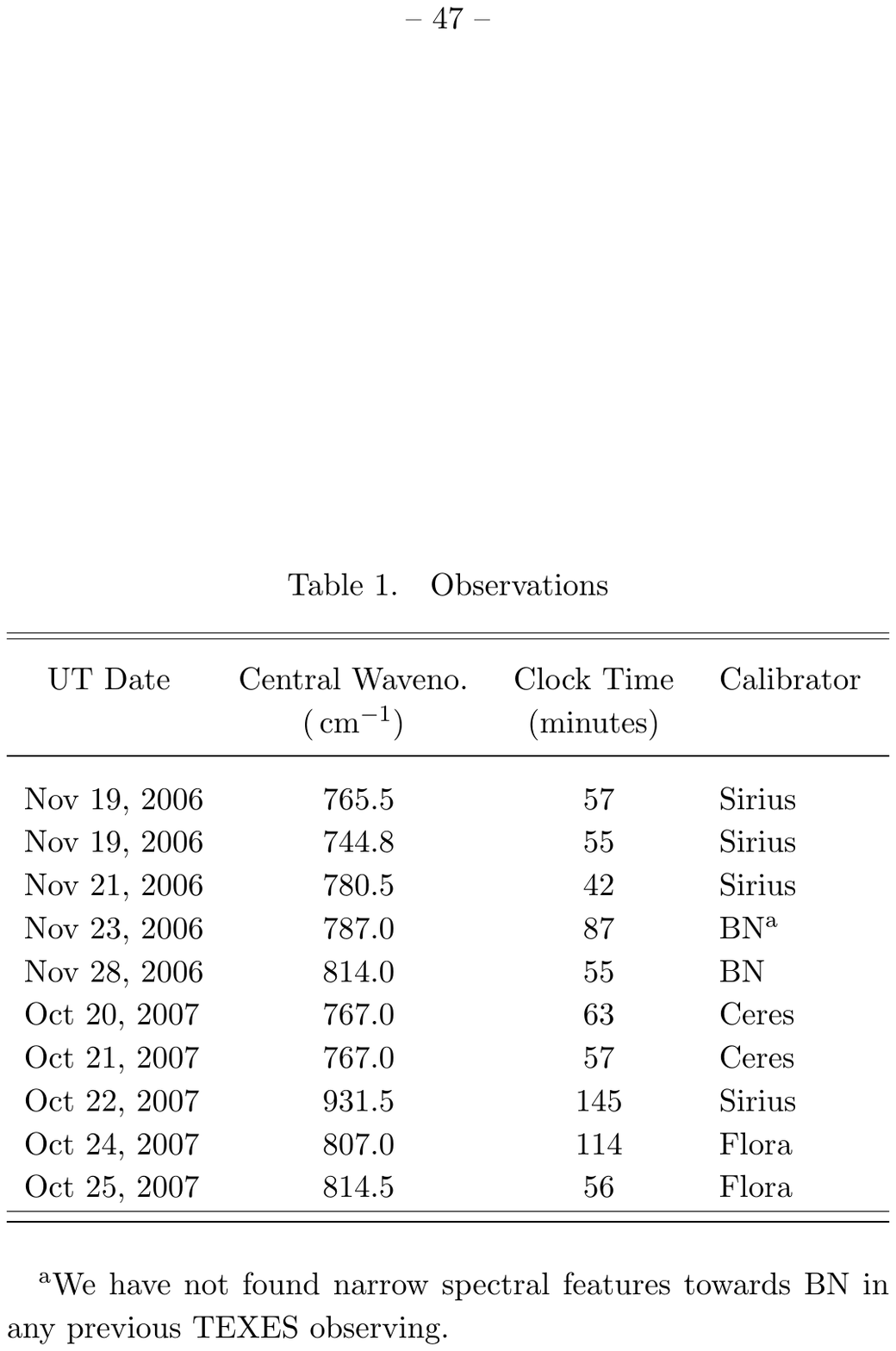}
\end{document}